\documentclass[%
reprint,
superscriptaddress,
nofootinbib,
showkeys,
amsmath,amssymb,
aps,
pra,
longbibliography,
floatfix,
nobalancelastpage
]{revtex4-2}

\usepackage{common}
\usepackage{xr}
\begin{document}

% ** Title page

\title{General Time-Dependent Configuration-Interaction Singles\\ \atomiccase: The Atomic Case}

\author{Stefanos Carlström\,\orcidlink{0000-0002-1230-4496}}%
\email{stefanos@mbi-berlin.de}
\email{stefanos.carlstrom@matfys.lth.se}
\affiliation{Max-Born-Institut, Max-Born-Straße 2A, 12489 Berlin, Germany}
\affiliation{Department of Physics, Lund University, Box 118, SE-221 00 Lund, Sweden}

\author{Mattias Bertolino\,\orcidlink{0000-0002-7430-7748}}%
\affiliation{Department of Physics, Lund University, Box 118, SE-221 00 Lund, Sweden}

\author{Jan Marcus Dahlström\,\orcidlink{0000-0002-5274-1009}}%
\affiliation{Department of Physics, Lund University, Box 118, SE-221 00 Lund, Sweden}

\author{Serguei Patchkovskii}%
\affiliation{Max-Born-Institut, Max-Born-Straße 2A, 12489 Berlin, Germany}

\date{\today}

\begin{abstract}
  We present a specialization of the grid-based implementation of the
  time-dependent configuration-interaction singles described in the
  preceding paper \cite{Carlstroem2022tdcisI} to the case of spherical
  symmetry. We describe the intricate time propagator in detail, and
  conclude with a few example calculations. Among these, of note are
  high-resolution photoelectron spectra in the vicinity of the Fano
  resonances in photoionization of neon, and spin-polarized
  photoelectrons from xenon, in agreement with recent experiments.
\end{abstract}

\keywords{Schrödinger equation, time-dependent
  configuration-interaction singles, photoelectron spectra,
  strong-field dynamics, spin--orbit dynamics, two-component
  quasi-relativistic}
\maketitle

% * Introduction
\section{Introduction}
This article describes the specialization of the general
time-dependent configuration-interaction singles presented in the
preceding article \cite[][hereinafter referred to as
\generalcase{}]{Carlstroem2022tdcisI} to the atomic case, taking
advantage of the spherical symmetry of the field-free
Hamiltonian. Where previously, the particle orbitals only had two
components, spin-up and spin-down (\(\alpha,\beta\equiv m_s=\pm1/2\)) respectively,
they are now expanded in spherical harmonics as well
(\(n\ell m_\ell s m_s\) basis), or directly in two-component spinor
spherical harmonics \cite[\(n\ell j m_j\) basis; see section~7.2
of][]{Varshalovich1988}. The two-component formulation allows us to
treat spin-dependent effects \emph{ab initio}, an important advance
beyond previous work
\cite{Rohringer2006,Rohringer2009PRA,Greenman2010PRA}. Since the
spin--angular algebra is fully analytic, the numerics are reduced to
coupled one-dimensional radial problems, which although constituting a
more compact basis than the three-dimensional Cartesian grids used in
the general case, also leads to comparatively more involved
expressions. The spectral properties of the matrix representations of
the various terms of the Hamiltonian also change, which requires
additional care when designing the time propagation scheme.

This article is arranged as follows: in
section~\ref{sec:atomic-structure}, the atomic structure problem is
briefly surveyed, section~\ref{sec:atomic-time-propagator}, which
constitutes the bulk of the paper, describes the details of the time
propagation scheme, and section~\ref{sec:example-calculations}
illustrates the implementation with some example
calculations. Finally, section~\ref{sec:atomic-conclusions} concludes
the paper. The same notation and conventions are used as detailed in
\ref{sec:conventions}. The atomic unit of time is
\(\SI{1}{jiffy}\approx\SI{24.2}{\atto\second}\), as introduced by
\textcite{Harriman1978}.

% * Atomic structure
\section{Atomic structure}
\label{sec:atomic-structure}

In contrast to the general case \generalcase{}, in the atomic case the
Hartree--Fock (HF) problem is solved on the same grid used to resolve
the particle orbitals \(\contket*{k},\contket*{l},...\). Furthermore,
we also require that the reference state is a solution to the HF
equations, instead of a general determinant. As a result, the matrix
representation of the \(\fock\) operator is diagonal in the space
spanned by occupied orbitals, which simplifies the equations of motion
(EOMs).

The Hamiltonian we consider is the following
\begin{equation}
  \label{eqn:full-hamiltonian}
  \begin{aligned}
    \Hamiltonian(t)
    &= \hamiltonian_i + \coulomb_{ij} + \laserinteraction[i](t) \\
    &=
      \frac{p_i^2}{2} +
      \frac{\ell(\ell+1)}{2r_i^2} +
      \nuclearpotential(r_i) +
      \complexabsorbingpotential(r_i) +
      \frac{1}{2r_{ij}} \\
    &\hphantom{=}+
      \begin{cases}
        \fieldamplitude[t]\cdot\vec{r}_i, & \textrm{(length gauge)}, \\
        \vectorpotential[t]\cdot\vec{p}_i + \identity_i\frac{A^2(t)}{2}, & \textrm{(velocity gauge)},
      \end{cases}
  \end{aligned}
\end{equation}
where \(\hamiltonian\) is the one-body Hamiltonian, \(\coulomb\) is
the two-body Coulomb electron--electron repulsion interaction [the term
\((2r_{ii})^{-1}\) is excluded from the summation], and
\(\laserinteraction(t)\) is the time-dependent interaction with an
external field. Each term will be described in more detail below. See
appendix~\ref{sec:quasimatrices} for a brief description of the
discretization of the radial problem.

% ** Relativistic effective core potentials
\subsection{Relativistic effective core potentials}
\label{sec:org7eb969f}
Although the EOMs \eqref{eqn:td-cis-eoms-atomic} are not
spin-restricted, they would yield the same result as a
one-component calculation, i.e.\ there would be no effect due to the
spin of the electrons. To implement spin--orbit coupling (and other
vector-relativistic effects), account for scalar-relativistic
effects, and at the same time reduce the number of electrons we
need to explicitly treat in the calculation, we replace
\(\nuclearpotential\) in \eqref{eqn:full-hamiltonian} by a
relativistic effective-core potential (RECP), which models the
nucleus and the core electrons according to
\begin{equation*}
  \pseudopotential(\vec{r}) =
  -\frac{Q}{r} + B_{\ell j}^k
  \exp(-\beta_{\ell j}^k r^2)
  \proj[\ell j],
\end{equation*}
where \(Q\) is the residual charge, \(\proj[\ell j]\) is a projector on
the spin--angular symmetry \(\ell j\), and \(B_{\ell j}^k\) and
\(\beta_{\ell j}^k\) are numeric coefficients found by fitting to excited
spectra computed using multiconfigurational Dirac--Fock all-electron
(AE) calculations. For a thorough introduction to RECPs, see e.g.\ the
review by \textcite{Dolg2011}. The use of RECPs allows us to
conveniently and accurately introduce the relativistic corrections
mentioned above into a two-component Schrödinger equation, instead of
having to resort to the four-component Dirac equation
\cite{Zapata2022}, which is more demanding computationally, and not
easily amenable for a grid formulation
\cite{Indelicato1993,Indelicato1995,Fischer2009}. In quantum chemistry
applications, methods employing RECPs are considered \emph{ab initio},
as they can be systematically improved
\cite{Walle1993,Pickard2001,Goddard2021}. It should be noted that
RECPs treat some of the relativistic terms, most notably the
electron--electron spin--orbit interaction, only in the mean-field
sense. Nonetheless, the errors introduced by this approximation are
generally small compared to the errors introduced by the CIS
\emph{Ansatz} \cite{Beck2009,Odoh2009}. In contrast, previous work
considering spin--orbit interaction on the TD-CIS level
\cite{Rohringer2009PRA,Pabst2012} consists of rotating from the
\(n\ell m_\ell s m_s\) basis to the \(n\ell j m_j\) basis and by introducing
the experimental ionization potentials for the \(J=1/2\) and \(J=3/2\)
(where \(J\) is the total angular momentum of the residual ion)
channels, respectively.

% * Time propagator
\section{Time propagator}
\label{sec:atomic-time-propagator}
The higher symmetry of the atomic case can be utilized when designing
the propagator, which is more efficient, but somewhat involved. The
chief reason for not using a polynomial approximation to the matrix
exponential, such as 4\textsuperscript{th} order Runge--Kutta (RK4) or
Krylov iterations, is the spectral range of the Hamiltonian which in
spherical coordinates is dominated by the centrifugal potential
\begin{equation}
  \label{eqn:centrifugal-potential}
  V_\ell(r) = \frac{\ell(\ell+1)}{2r^2},
\end{equation}
the highest eigenvalue of which is on the order of
\(\ellmax^2/\rmin^2\). This severely limits the largest time step that
can be taken by the propagator. Instead, we opt for a second-order
palindromic Strang splitting \cite{Strang1968} (cf.\ the symmetric
Baker--Campbell--Hausdorff formula) of the propagator
\begin{widetext}
  \begin{equation*}
    \begin{aligned}
      \exp\left[
      \timeord
      \int_0^{\timestep}\diff{t}
      \mat{M}(t)
      \right] =
      ...\ce^{\timestep\mat{C}/2}
      \ce^{\timestep\mat{B}/2}
      \ce^{\timestep\mat{A}}
      \ce^{\timestep\mat{B}/2}
      \ce^{\timestep\mat{C}/2}...
      +
      \Ordo\{\timestep^3(\comm{\mat{A}}{\mat{B}} + \comm{\mat{A}}{\mat{C}} + \comm{\mat{B}}{\mat{C}} + ...)\},
    \end{aligned}
  \end{equation*}
\end{widetext}
where \(\frac{1}{\timestep}\int_0^{\timestep}\diff{t}\mat{M}(t) =
\mat{A}+\mat{B}+\mat{C}...\) is the matrix representation of the
Hamiltonian \(-\im\Hamiltonian(t)\) integrated over the time step
\(\timestep\), and \(\timeord\) is the time-ordering operator. This
splitting lets us tailor a propagator for each part of the
Hamiltonian which may have vastly different spectral and spatial
properties, e.g.\ the centrifugal potential
\eqref{eqn:centrifugal-potential} can trivially be exponentiated
exactly, circumventing the issues of large spectral radius which is
problematic for polynomial approximations. In contrast, a simple RK4
propagator is used for the Coulomb interaction, whose spectral
radius is rather limited, but which is costly to evaluate. The use
of RK4 makes the propagator only conditionally stable, however this
has not been found to be a problem in practice. Finally, since the
overall splitting is second-order in time, it is enough to integrate
the time-dependent terms of the Hamiltonian to the same order, for
instance via evaluation at the centre of the time step.

Which terms appear in the splitting depends on the particular
system, but the general structure is
\begin{equation*}
  \propU =
  \markterm{1}{\revexpprop[\timestep \dipolemat/2]}
  \markterm{2}{\expprop[\timestep \mat{A}]}
  \markterm{3}{\expprop[\timestep \dipolemat/2]} +
  \Ordo\{\timestep^3[...]\},
\end{equation*}
where \(\atomicmat\) contains the field-free Hamiltonian,
\(\dipolemat\) the dipole interaction, and \(\revexpprop\) indicates
that any possible subsplitting is applied in reverse (to preserve
unitarity to \(\Ordo\{\timestep^3\}\)). In the most complicated
case, with dipole couplings between the occupied orbitals (e.g.\ in
neon), and spin--orbit interaction, the full propagator reads
\begin{widetext}
  \begin{equation}
    \label{eqn:full-propagator}
    \propU =
    \markterm{1}{
      \revexpprop[\timestep \dipolematss/2]
      \revexpprop[\timestep \dipolematvv/2]
      \revexpprop[\timestep \dipolematsv/2]
    }\;
    \markterm{2}{
      \revexpprop[\timestep \atomicmat_{\textrm{1b}}/2]
      \revexpprop[\timestep \atomicmat_{\textrm{s--o}}/2]
      \expprop[\timestep \atomicmat_{\textrm{2b}}]
      \expprop[\timestep \atomicmat_{\textrm{s--o}}/2]
      \expprop[\timestep \atomicmat_{\textrm{1b}}/2]
    }\;
    \markterm{3}{
      \expprop[\timestep \dipolematsv/2]
      \expprop[\timestep \dipolematvv/2]
      \expprop[\timestep \dipolematss/2]
    } +
    \Ordo\{\timestep^3[...]\},
  \end{equation}
\end{widetext}
whereas in the non-relativistic single-active electron (SAE) case
(e.g.\ hydrogen), the propagator reduces to
\begin{equation*}
  \propU =
  \markterm{1}{\revexpprop[\timestep \dipolematvv/2]}
  \markterm{2}{\expprop[\timestep \atomicmat_{\textrm{1b}}]}
  \markterm{3}{\expprop[\timestep \dipolematvv/2]} +
  \Ordo\{\timestep^3\comm{\atomicmat_{\textrm{1b}}}{\dipolematvv}\}.
\end{equation*}
The various subterms will be enumerated and described in the
following sub-sections.

We note that the above scheme is similar in spirit to those of
\textcite{Sato2016,Teramura2019}, but differs in details. The latter
has higher convergence order than the present work, but relies on the
\(\varphi_k\) matrix functions \cite{Hochbruck2010} which can be numerically
delicate to implement. Additionally, our scheme is rather different
from earlier atomic TD-CIS implementations described in the
literature: \textcite{Rohringer2006} used RK4 since they only
considered 1D systems where the centrifugal potential does not appear;
\textcite{Rohringer2009PRA,Greenman2010PRA} instead represent their
wavefunctions in the basis of the singly excited Slater determinants
(which yields block-dense dipole and Coulomb matrices), using a
second-order differencing propagator. Compared to these
implementations, our scheme can handle comparatively large time steps.

% ** Propagation on a submanifold
\subsection{Propagation on a submanifold}
\label{sec:propagation-on-submanifold}
\begin{figure*}[htb]
  \centering
  \includegraphics{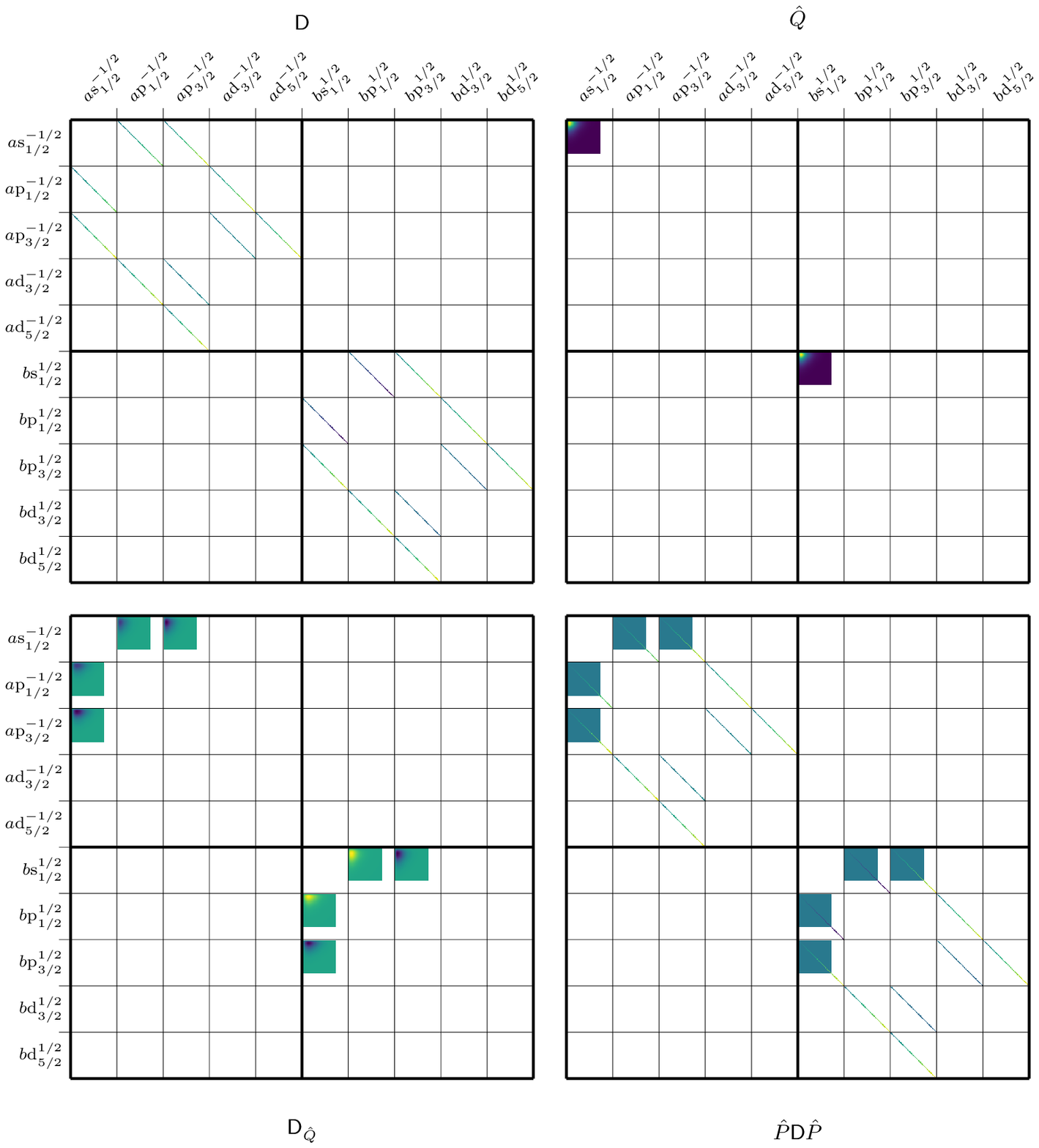}
  \caption{\label{fig:projected-dipole-sparsity}Sparsity pattern of
    dipole operator \(\dipolemat\), in the case of helium in the
    \(n\ell jm_j\) basis (\(\ell\in\{\conf{s,p,d}\}\)) and linear polarization
    (\(\implies\Delta m_j=0\)). Each block corresponds to coupling between
    a pair of particle orbitals, which are labelled by their particles
    \(\ell_j^{m_j}\), and their holes
    \(a\iff\slaterhole{\conf{1s}_{1/2}^{-1/2}}\) and
    \(b\iff\slaterhole{\conf{1s}_{1/2}^{1/2}}\). See main text for
    discussion.}
\end{figure*}

\begin{figure}[htb]
  \centering
  \includegraphics{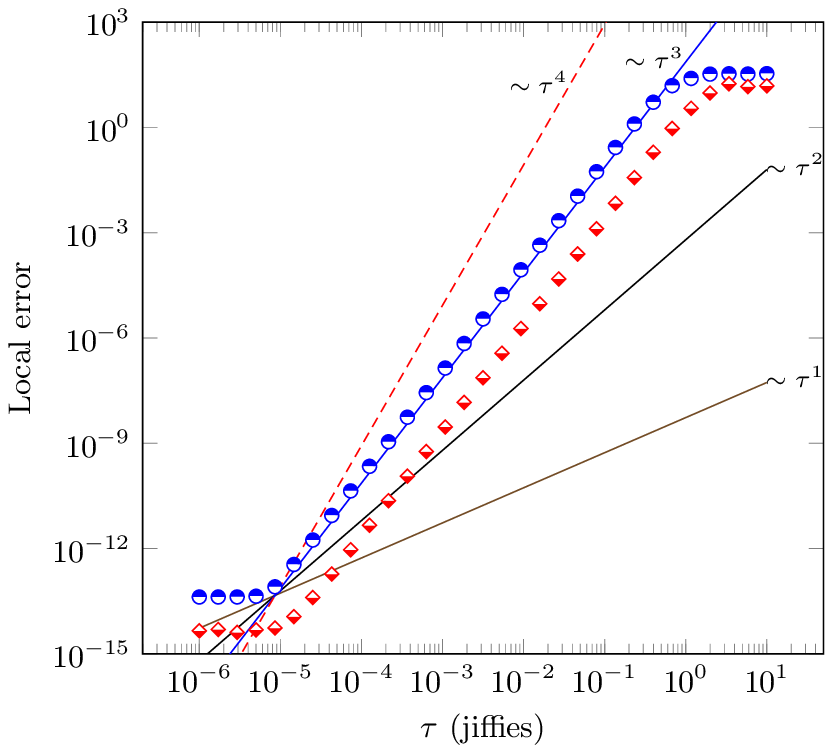}
  \caption{\label{fig:projected-dipole-splitting-error}Performance of
    the Strang splitting of the projected dipole illustrated in
    Figure~\ref{fig:projected-dipole-sparsity}; the blue circles
    indicate the local error
    \(||{\exp}(-\im\timestep\proj\dipolemat\proj)-{\exp}(-\im\timestep\rejector{\dipolemat}/2){\exp}(-\im\timestep\dipolemat){\exp}(-\im\timestep\rejector{\dipolemat}/2)||_2\)
    as a function of the time step
    \(\timestep\). \({\exp}(-\im\timestep\proj\dipolemat\proj)\) is
    computed exactly via diagonalization,
    \({\exp}(-\im\timestep\dipolemat)\) via Givens rotations (see
    section~\ref{sec:dipole-propagation} below), and
    \({\exp}(-\im\timestep\rejector{\dipolemat}/2)\) using RK4. The
    straight lines show that the local error indeed is of
    \(\Ordo\{\timestep^3\}\), as expected. Another measure of the
    accuracy shown as red diamonds, is
    \(||\rej{\exp}(-\im\timestep\rejector{\dipolemat}/2){\exp}(-\im\timestep\dipolemat){\exp}(-\im\timestep\rejector{\dipolemat}/2)\proj||_2\),
    the norm of the \emph{rejection of the split-propagator itself},
    i.e.\ how much the split-propagator populates
    \(\perp\submanifold\), i.e.\ the orthogonal complement of the
    submanifold. This measure also behaves as
    \(\Ordo\{\timestep^3\}\).}
\end{figure}

When propagating the EOMs \eqref{eqn:td-cis-eoms-atomic}, we are
solving a partial differential equation on a submanifold
\(\submanifold\subset\hilbertspace\), due to the constraint that
the particle orbitals must at all times remain orthogonal to the
occupied orbitals; in general,
\(\Hamiltonian\Psi\not\in\submanifold\). This means that instead of
computing the matrix exponential \(\propU=\exp(\mat{A})\), as we
would ordinarily do if the solution was allowed to occupy any part
of the Hilbert space \(\hilbertspace\), we need to compute its
projected counterpart
\begin{equation*}
  \begin{aligned}
    \propU_p
    &=
      {\exp}(\proj\mat{A}\proj),
    &
      \proj
    &=
      \identity - \rej,
    &
      \rej
    &=
      \ketbra{i}{i}.
  \end{aligned}
\end{equation*}
Normally, for a matrix \(\mat{A}\) with a similarity transform
\(\mat{S}\), we have
\begin{equation*}
  \exp(\mat{S}\mat{A}\mat{S}^{-1}) =
  \mat{S}\exp(\mat{A})\mat{S}^{-1},
\end{equation*}
which is most easily proved using the \(k\)th term of the Taylor
expansion of the exponential function:
\begin{equation*}
  \frac{(\mat{S}\mat{A}\mat{S}^{-1})^k}{k!} =
  \mat{S}\frac{\mat{A}^k}{k!}\mat{S}^{-1}.
\end{equation*}
We cannot use this relation in the present case since the
projectors are idempotent:
\begin{equation*}
  \proj^2 = \proj \implies
  {\exp}(\proj\mat{A}\proj) \neq
  \proj\exp(\mat{A})\proj,
\end{equation*}
and making this approximation would reduce convergence of the time
propagator to first order.

Furthermore, since \(\proj\) is spatially dense, any sparsity
pattern of \(\mat{A}\) that we hoped to benefit from seems
lost. However, we can use the fact that the occupied orbitals
\(\ket{i}\) from which the projectors are constructed are spatially
confined to the extents of the HF reference, together with the
splitting:
\begin{equation}
  \label{eqn:rejector-splitting}
  \proj\mat{A}\proj =
  (\identity-\rej)
  \mat{A}
  (\identity-\rej) =
  \mat{A} +
  \underbrace{(-\rej\mat{A}-\mat{A}\rej +
    \rej\mat{A}\rej)}_{\defd \rejector{\mat{A}}}.
\end{equation}
Since \(\rej\) is the projector onto the space of occupied
orbitals, it is limited in extent, i.e.~it has compact support,
which in turn means that \(\rejector{\mat{A}}\) has compact
support. This is crucial, because it means we can precompute the
exponential of \(\rejector{\mat{A}}\) via exact diagonalization,
and we can approximate \(\exp(\timestep\mat{A})\) using any method
of our choosing.

Finally, \(\proj\Psi\) projects \(\Psi\) onto the submanifold
\(\submanifold\), and conversely \(\rej\Psi\) is then the
\emph{rejection}. We thus term \(\rejector{\mat{A}}\) the
\emph{rejector} of \(\mat{A}\). To illustrate the efficacy of the
rejector splitting \eqref{eqn:rejector-splitting}, we consider the
dipole interaction (see \ref{sec:dipole-propagation} below); in
Figure~\ref{fig:projected-dipole-sparsity}, the sparsity patterns of
the dipole interaction \(\dipolemat\), and its projectors and
rejectors are shown. Since in finite-differences, potentials are
represented by diagonal matrices, non-zero blocks of \(\dipolemat\)
only have entries on their diagonals. The other matrices are, as
labelled, the sparsity patterns of the projection \(\rej\) onto the
occupied orbitals, which is non-zero only for those to particle
orbitals that are in the same symmetries as the two occupied orbitals
of the HF reference; the rejector
\(\rejector{D}=-\rej\dipolemat-\dipolemat\rej+\rej\dipolemat\rej\),
which is spatially compact since it 1) only couples a few particle
orbitals, and 2) radially only encompasses the extents of the HF
reference, thereby not filling the whole matrix block; and the
projected dipole \(\proj\dipolemat\proj\), which is the operator we
wish to approximate the exponential of, but whose sparsity pattern is
unfavourable to any approximations beyond polynomial methods. In this
illustrative example, the HF problem is solved on the radial interval
\SIrange{0}{7}{Bohr}, whereas the dipole is computed on the interval
\SIrange{0}{10}{Bohr}; in a more realistic scenario, the dipole
operator is required for \(r\GG\SI{10}{Bohr}\), increasing the
numerical utility of this operator splitting.

When approximating \({\exp}(-\im\timestep\proj\dipolemat\proj)\) by
\({\exp}(-\im\timestep\rejector{\dipolemat}/2){\exp}(-\im\timestep\dipolemat){\exp}(-\im\timestep\rejector{\dipolemat}/2)\),
and varying the time step \(\timestep\), the local error is cubic in
\(\timestep\), which leads to a second-order propagator overall (see
Figure~\ref{fig:projected-dipole-splitting-error}).

% ** One-body atomic Hamiltonian
\subsection{One-body atomic Hamiltonian}
The one-body part of the atomic Hamiltonian, labelled
\(\atomicmat_{\textrm{1b}}\) in \eqref{eqn:full-propagator},
contains the kinetic energy and one-body potential energy of the
electron:
\begin{equation}
  \label{eqn:one-body-hamiltonian}
  \hamiltonian \defd \frac{p^2}{2} +
  \frac{\ell(\ell+1)}{2r^2} +
  \nuclearpotential(r) +
  \complexabsorbingpotential(r),
\end{equation}
where the \emph{complex absorbing potential} (CAP)
\(\complexabsorbingpotential(r)\) is usually taken to be that of
\textcite{Manolopoulos2002}. Since in finite-differences, the matrix
representation of \eqref{eqn:one-body-hamiltonian} is a tridiagonal
matrix, we use the Crank--Nicolson method to approximate the matrix
exponential. The orbitally diagonal part of the direct interaction
\(\direct[ii]\) (see next section), i.e.\ the Hartree potential, is
although formally a two-body operator, effectively a one-body
potential and as such a diagonal matrix, which we exponentiate
together with \(\hamiltonian\).

Acting with \(\hamiltonian\) on a particle orbital
\(\contket*{k}\) can take us out of the correct submanifold, as
described in the preceding section. We therefore precompute
\({\exp}(-\im \timestep \rejector{\hamiltonian}/2)\), where
\(\rejector{\hamiltonian}\) is defined in
\eqref{eqn:rejector-splitting}, and place this on either side of
the Crank--Nicolson propagator for \(\hamiltonian\).

Even though adding a CAP may be seen as a pragmatic approach to avoid
unwanted reflections, they are systematically improvable to the point
that they are formally equivalent to exterior complex scaling
\cite{Riss1993,Riss1998,Moiseyev1998a,Moiseyev1998,Santra2002,Scrinzi2010PRA,Moiseyev2011}. TD-CIS
with the addition of a CAP can thus still be considered \emph{ab
  initio} \cite{Greenman2010PRA}.

% ** Coulomb interaction
\subsection{Coulomb interaction}
The Coulomb interaction, labelled \(\atomicmat_{\textrm{2b}}\) in
\eqref{eqn:full-propagator}, enters the TD-CIS EOMs
\eqref{eqn:td-cis-eoms-atomic} through the Fock operator
\(\fock=\hamiltonian+\direct[ii]-\exchange[ii]\) which appears on the
orbital diagonal, as well as the configuration interaction term
\(\direct[lk]-\exchange[lk]\) which couples the different
particle--hole channels. Since the occupied orbitals remain fixed in
the TD-CIS \emph{Ansatz}, both \(\direct\) and \(\exchange\) remain
formally time-independent. However, whereas \(\direct\) is only
non-local in the spin--angular coordinates (and thus radially
diagonal), \(\exchange\) is also radially non-local, preventing its
storage as a structured matrix. On the other hand, \(\exchange\) has
radially compact support. To see this, we rewrite the Coulomb
interaction \eqref{eqn:mulliken} in spherical coordinates \cite[see
Equation~(5.17.9) of][]{Varshalovich1988}:
\begin{equation*}
  \twobody{ab}{cd} =
  \sum_k
  C_{abcd}^k
  \int_0^\infty\diff{r_1}
  \conj{P_a}(r_1)
  \frac{Y_{bd}^k(r_1)}{r_1}
  P_c(r_1),
\end{equation*}
where \(C_{abcd}^k\) are Clebsch--Gordan coefficients, and the
\(k\)th multipole of the repulsion potential formed by the orbitals
\(\orbital{i},\orbital{j}\) is given by
\begin{equation*}
  \begin{aligned}
    Y^k_{ij}(r_1)
    &\defd
      r_1\int_0^\infty
      \diff{r_2}
      \frac{r_<^k}{r_>^{k+1}}
      \conj{P_i}(r_2)
      P_j(r_2) \\
    &=
      \int_0^{r_1}
      \diff{r_2}
      \left(\frac{r_2}{r_1}\right)^k
      \conj{P_i}(r_2)
      P_j(r_2) \\
    &\hphantom{=}+
      \int_{r_1}^\infty
      \diff{r_2}
      \left(\frac{r_1}{r_2}\right)^{k+1}
      \conj{P_i}(r_2)
      P_j(r_2),
  \end{aligned}
\end{equation*}
and \(P_i(r),P_j(r)\) are the radial components of the
orbitals. These potentials are found by solving Poisson's problem
\cite{Fischer1977,Fischer1990,McCurdy2004,McCurdy2004a}, with the
mutual charge density \(\rho_{ij}(r)=\conj{P_i}(r)P_j(r)\) as the
inhomogeneous source term. In the TD-CIS \emph{Ansatz}, this charge
density is formed from one (in case of \(\exchange\)) or two (in
case of \(\direct\)) occupied orbitals, which means we only have to
solve Poisson's problem on the radial extent of the HF
reference. However, we then need to add in a homogeneous
contribution as well, that accounts for the long-range behaviour;
this is only trivial to do in the spherically symmetric case and is
an important optimization over the general case, where instead
Poisson's problem has to be solved over the entire domain (however,
in that case an asymptotic multipole solution may be used as the
initial guess, speeding up convergence of the solution).

From this argument, we see that \(Y_{ij}^k(r)\) for the direct
interaction \(\direct\) will be formed from two occupied orbitals
\(\ket{i},\ket{j}\), and will thus be time-independent and radially
diagonal (i.e.\ represented by a diagonal matrix in
finite-differences), however, it will extend over the whole
computational domain. In contrast, to compute the exchange
interaction \(\exchange\), the mutual charge density is formed from
an occupied orbital \(\ket{i}\) and the time-dependent particle
orbital \(\contket*{l}\) which \(\exchange\) acts on, which means
the resulting potential is radially non-local. The mutual charge
density will have zero trace, and \(Y_{i\cont{l}}^k(r)\) for
\(\exchange\) will decay as at least \(r^{-2}\). Additionally, it
is subsequently applied to an occupied orbital, which decays as
\({\exp}(-\sqrt{2\abs{\orbitalenergy{i}}}r)\). Thus the \(\exchange\)
operator is radially localized to the HF reference. Acting with
\(\direct\) on the wavefunction thus amounts to multiplying with
precomputed radially diagonal matrices, and acting with
\(\exchange\) amounts to solving a radially localized Poisson
problem. These operations are the limiting factors of the time
propagator.

Since the Coulomb interaction is of limited spectral range, a
polynomial approximation to the matrix exponential that has a fixed
number of matrix--vector products per step (such as RK4), is
entirely satisfactory. Maintaining orthogonality of the particle
orbitals with respect to the occupied orbitals is trivial by
projecting out the latter after each RK4 stage. This procedure is
similar to the approach taken by \textcite{Sato2016}.

% ** Dipole interaction
\subsection{Dipole interaction}
\label{sec:dipole-propagation}
The interaction with the external laser field is treated in the
dipole approximation, where the two most common choices for the
interaction operator are
\begin{equation*}
  \laserinteraction(t)
  =
  \begin{cases}
    \fieldamplitude[t]\cdot\vec{r}, & \textrm{(length gauge)}, \\
    \vectorpotential[t]\cdot\vec{p} + \frac{A^2(t)}{2}, & \textrm{(velocity gauge)}.
  \end{cases}
\end{equation*}

Although the TD-CIS \emph{Ansatz} with frozen core orbitals is gauge
variant \cite{Wolfsberg1955,Kobe1979,Ishikawa2015}, we have
implemented dipole interaction for both gauges; all results
presented in the present work are however computed in the length
gauge.

There are three terms we need to consider, the \emph{source--virtual}
dipole interaction \(\matrixel*{k}{\laserinteraction}{\cont{k}}\),
the \emph{virtual--virtual} dipole interaction
\(\matrixel*{\cont{k}}{\laserinteraction}{\cont{k}}\), and the
\emph{source--source} dipole interaction
\(\matrixel{k}{\laserinteraction}{l}\) (may be absent in some
systems).

% *** Source--virtual dipole interaction
\subsubsection{Source--virtual dipole interaction}
This interaction, labelled \(\dipolematsv\) in
\eqref{eqn:full-propagator}, corresponds to the sub-EOMs
\begin{equation}
  \label{eqn:source-virtual-dipole-interaction}
  \begin{aligned}
    \imdt c_0
    &=
      \markterm{1}{\matrixel*{k}{\laserinteraction}{\cont{k}}}, \\
    \imdt\contket*{k}
    &=
      \markterm{1}{c_0\laserinteraction\ket{k}} -
      \lagrange{\cont{k}i}\ket{i},
  \end{aligned}
\end{equation}
which we call the source--virtual dipole interaction, since the
occupied orbital \(\ket{k}\) constitutes a source term for the
particle orbital \(\contket*{k}\), which in turn is a linear
combination of virtual orbitals. We can rewrite the
EOMs~\eqref{eqn:source-virtual-dipole-interaction} on matrix form
\begin{equation}
  \label{eqn:source-virtual-dipole-interaction-matrix}
  \tag{\ref{eqn:source-virtual-dipole-interaction}*}
  \imdt q =
  \underbrace{\bamat{c|c}{0&\bra{m}\\\hline\ket{m}&0}}_{\defd A}
  q, \quad
  q\defd\bmat{c_0\\\contket*{m}},
\end{equation}
where the matrix \(A\) is non-zero only in the first column and
row, respectively, and
\begin{equation*}
  \ket{m}\defd\bmat{\proj\laserinteraction\ket{k}\\\proj\laserinteraction\ket{l}\\\vdots},
  \quad
  \contket*{m}\defd\bmat{\contket*{k}\\\contket*{l}\\\vdots}.
\end{equation*}
The projector \(\proj\) in \(\ket{m}\) ensures orthogonality of
the particle orbitals to the occupied orbitals, after
applying the source--virtual dipole interaction. Since the laser
interaction \(\laserinteraction=\fieldamplitude[t]\cdot\vec{r}\)
is time-dependent, \(\ket{m}\) has to be recomputed every time
step. However, the projected polarized source orbitals
\begin{equation*}
  \bmat{
    \proj
    \operator{d}\ket{k}\\
    \proj
    \operator{d}\ket{l}\\
    \vdots}, \quad
  \operator{d}=\operator{x},\operator{y},\operator{z},
\end{equation*}
can be precomputed, and linearly combined with the time-dependent
field components \(\fieldcomponent[_d](t)\).

Having formed the matrix
EOMs~\eqref{eqn:source-virtual-dipole-interaction-matrix}, we can
solve it exactly, if we can form the singular-value decomposition
(SVD) of \(\mat{A}=\mat{S}\tilde{\mat{A}}\adjoint{\mat{S}}\) (this
choice is possible if \(\mat{A}\) is Hermitian):
\begin{equation}
  \label{eqn:sv-svd-exp}
  \exp(\mu A)q =
  \{{S}[{\exp}(\mu\tilde{A})-\identity]\adjoint{S}+\identity\}q.
\end{equation}
For the \(\mat{A}\) we have in TD-CIS, the SVD that decomposes
\(\mat{}A\) is given by
\begin{equation*}
  \begin{aligned}
    &\tilde{A}
      \defd
      \bmat{s&0\\0&-s},
                    \quad s\defd\abs{\braket{m}{m}} \\
    &\implies
      {\exp}(\mu\tilde{A}) - \identity
      \equiv
      \bmat{\ce^{\mu s}-1&0\\0&\ce^{-\mu s}-1},
  \end{aligned}
\end{equation*}
and the left-singular vectors are given by
\begin{equation*}
  \mat{S} \defd
  \frac{1}{\sqrt{2}s}
  \bmat{s&-s\\
    \ket{m}&\ket{m}}.
\end{equation*}
One time step with the source--virtual dipole interaction can thus
be accomplished by
\begin{equation}
  \tag{\ref{eqn:sv-svd-exp}*}
  \begin{aligned}
    &\bmat{c_0\\\contket*{m}} \leftarrow
    \bmat{c_0\\\contket*{m}} \\
    &+
      \frac{1}{2s^2}
      \bmat{s&-s\\
    \ket{m}&\ket{m}}
             \bmat{\ce^{\mu s}-1&0\\0&\ce^{-\mu s}-1}
                                     \bmat{c_0s + \braket{m}{\cont{m}}\\
    -c_0s + \braket{m}{\cont{m}}}.
  \end{aligned}
\end{equation}
The advantage of this formulation, is that the complexity of the
matrix exponential reduces to linear in the size of the compact HF
support. We note that although our EOMs
\eqref{eqn:td-cis-eoms-atomic} are non-Hermitian due to the
presence of a CAP, we still have that
\(\adjoint{\ket{m}}\equiv\bra{m}\), since \(\ket{m}\) is formed as
a linear combination of projected polarized source orbitals, and
thus of the same radial extent as the HF reference, where the CAP
is identically zero.

% *** Virtual--virtual dipole interaction
\subsubsection{Virtual--virtual dipole interaction}
This interaction, labelled \(\dipolematvv\) in
\eqref{eqn:full-propagator}, is almost the same as in the SAE case,
and therefore it is implemented analogously (see
appendix~\ref{sec:givens-rotations} and
e.g.~\textcite{Muller1999LP,Schafer2009,Patchkovskii2016}), with added
complication in those partial waves which share spin--angular quantum
numbers with the occupied orbitals of the HF reference; for those we
employ the ideas detailed in
section~\ref{sec:propagation-on-submanifold}. Specifically, for the
propagation of the rejector of the dipole, \(\rejector{\dipolemat}\),
we use RK4 to approximate the matrix exponential.

For systems treated in the \(n\ell j m_j\) basis, we temporarily
change to the \(n\ell m_\ell s m_s\) basis via a unitary
transformation (built from Clebsch--Gordan coefficients), since the
dipole interaction is sparser in that representation.

% *** Source--source dipole interaction \& spin--orbit interaction
\subsubsection{Source--source dipole interaction \& spin--orbit interaction}
For systems where there are dipole moments between the occupied
orbitals (e.g.\ neon), the dipole interaction can trigger
transitions between the channels; we call this source--source
dipole interaction, and it is labelled \(\dipolematss\) in
\eqref{eqn:full-propagator}. For each pair \(\ket{k},\ket{l}\) of
orbitals which have a non-zero dipole moment, the corresponding
pair of particle orbitals \(\contket*{k},\contket*{l}\) are mixed
using a Givens rotation (cf.\ appendix~\ref{sec:givens-rotations}) where
the rotation angle is \(a=\fieldamplitude[t]\cdot\vec{d}_{kl}\),
and the dipole moment
\(\vec{d}_{kl}\defd\matrixel{\orbital{k}}{\vec{r}}{\orbital{l}}\)
is precomputed. Exactly the same approach is taken in the case of
spin--orbit interaction between occupied orbitals [labelled
\(\atomicmat_{\textrm{s--o}}\) in \eqref{eqn:full-propagator},
e.g.\ \(\conf{4p}\) and \(\conf{5p}\) in xenon], with the only
difference that the spin--orbit interaction is time-independent.

% *** General polarization
\subsubsection{General polarization}
The Cartesian operators \(x\), \(y\), \(z\) commute, however in a
truncated spherical basis only approximately so, e.g.\ the
commutator \(\comm{z}{x}\) is non-zero only in the highest
considered \(\ell\) channel. We can thus safely use the splitting
\(\exp(\mu\fieldamplitude\cdot\vec{r})=
\exp(\mu\fieldcomponent[_x]x) \exp(\mu\fieldcomponent[_y]y)
\exp(\mu\fieldcomponent[_z]z)\), as long as the population in the
highest \(\ell\) channel is negligible. We stay in the lab frame,
i.e.\ we do not rotate the wavefunction as done by
\textcite{Muller1999LP,Patchkovskii2016}, since although that would
potentially be more efficient, it would require the rotation of
the occupied orbitals \(\ket{i},\ket{j}\) (and hence the
potentials \(\direct[ij]\) and \(\exchange[ij]\)), in addition to
the particle orbitals \(\contket*{i},\contket*{j}\) (which are
the analogues of the wavefunctions in the SAE case). Although this
is technically possible, the numerical implementation is
non-trivial and error-prone.

% * Example calculations
\section{Example calculations}
\label{sec:example-calculations} Unless otherwise specified, the
calculations below use \emph{truncated} Gaussian pulse envelopes
\cite{Patchkovskii2016} for the vector potential:
\begin{equation}
  \begin{aligned}
    \label{eqn:truncated-gaussian}
    \vec{A}(t) &= \vec{A}_0
                 \exp[-\alpha f(t)]
                 \sin(\omega t + \phi), \\
    f(t) &\defd \begin{cases}
                  \abs{t}^2 & \abs{t} \leq \toff,\\
                  \abs{\toff +
                  \frac{2t_{\textrm{mo}}}{\pi}
                  \tan\left(
                  \frac{\pi}{2}\frac{\abs{t}-\toff}{t_{\textrm{mo}}}
                  \right)
                  }^2,
                            & \toff < \abs{t} \leq \tmax,\\
                  +\infty, & \textrm{else},
                \end{cases}
  \end{aligned}
\end{equation}
where the helper function \(f(t)\) ensures a smooth turn-off of the
field starting at \(\toff\) and finishing at \(\tmax\)
(\(t_{\textrm{mo}}\defd\tmax-\toff\)). The parameter \(\alpha\) is
determined such that the full-width half-maximum \(\pulseduration\) of
the intensity envelope is the desired pulse duration; since
\(\vec{F}(t)\defd-\partial_t\vec{A}(t)\) the map from
\(\pulseduration\) to \(\alpha\) is in general dependent on the carrier
angular frequency \(\omega\), but in the long-pulse limit
\(\alpha\to 2\ln2/\pulseduration^2\). Typically, we choose
\(\toff=4\sigma\) and \(\tmax=6\sigma\), where the standard deviation of the
intensity envelope is given by
\(\sigma=\pulseduration/(2\sqrt{2\ln2})\). The main benefit of this pulse
shape is the suppression of side-lobes in the spectrum (which can lead
to e.g.\ overestimation of one-photon cross-sections), while still
maintaining a minimal time--bandwidth product. An example spectrum for
a two-colour field with the envelope \eqref{eqn:truncated-gaussian} is
shown in Figure~\ref{fig:neon-spectrum}~(b).

For the example calculations presented below, the radial grid
employed is smoothly approaching a uniform grid according to the
formula \cite{Krause1999TJoPCA}:
\begin{equation}
  \label{eqn:radial-grid}
  r_j = r_{j-1} + \rhomin + (1 - \ce^{-\alpha
    r_{j-1}})(\rhomax-\rhomin),
\end{equation}
with the first grid point at \(r_1=\rhomin/2\). This yields an
approximately log--lin behaviour with a dense grid close to the origin
where the bound orbitals exhibit a very oscillatory behaviour. The
asymptotic grid spacing should be chosen to fulfil the Nyquist
sampling theorem for the highest momentum desired:
\(\rhomax\lesssim(2\maxq{k})^{-1}=(8\kinengmax)^{-1/2}\), where
\(\kinengmax=\SI{4}{\hartree}\), unless otherwise specified. The
specific grid parameters are given in Table~\ref{tab:grid-parameters}.
\begin{table}[htbp]
  \caption{\label{tab:grid-parameters}Radial grid parameters used for
    the example calculations.}
  \centering
  \begin{ruledtabular}
    \begin{tabular}{lrrrrl}
      & \ch{H}\Bstrut & \ch{He} & \ch{Ne} AE & \ch{Ne} RECP & \ch{Xe} RECP\\
      \hline
      \(\rhomin\)\Tstrut (Bohr) & 0.3 & 0.15 & 0.1 & 0.125 & 3/26\\
      \(\alpha\) & 0.1 & 0.1 & 0.1 & 0.3 & 0.3
    \end{tabular}
  \end{ruledtabular}
\end{table}

The number of partial waves scales quadratically with the maximum
orbital angular momentum \(\ellmax\), and linearly with number of
channels \(n_c\):
\begin{equation}
  \label{eqn:num-pws}
  n_p = 2(\ellmax+1)^2n_c,
\end{equation}
where the factor of \(2\) comes from the spin of the excited
electron. In the case of linear polarization along \(z\), where
\(\Delta m_\ell=0\) (\(\Delta m_j=0\)), this reduces to
\begin{equation}
  \label{eqn:num-pws-linear}
  n_p < \kappa(\ellmax+1)n_c,\quad
  \kappa=
  \begin{cases}
    1, & n\ell m_\ell s m_s,\\
    2, & n\ell j m_j,
  \end{cases}
\end{equation}
where we can only give an upper bound since \(\ellmin\) depends on the
ionization channel (i.e.\ \(\ell\ge\abs{m_\ell}\)).

% ** Static polarizability
\subsection{Static polarizability}
\begin{figure*}[htb]
  \centering
  \includegraphics{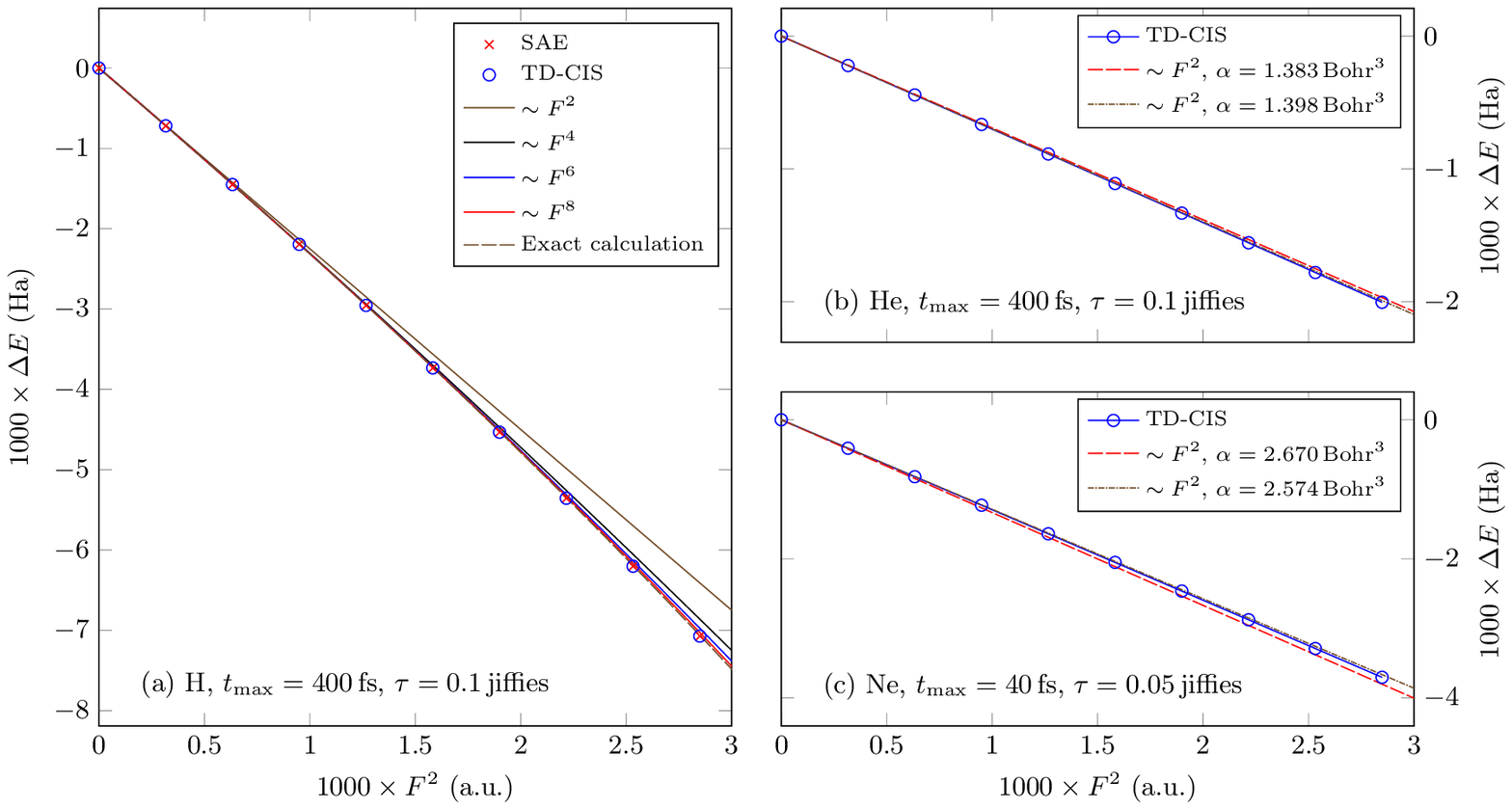}
  \caption{\label{fig:static-polarizability-hydrogen}Static
    polarizability of (a) hydrogen, (b) helium, and (c) neon, as a
    function of static field strength \(F\). Also plotted are
    expansions in powers of \(F^2\) with coefficients taken from
    theory (hydrogen) or experiment (helium and neon). For hydrogen,
    adding more terms to the expansion in \(F^2\) approaches the
    results as computed by SAE and TD-CIS. Also shown are exact
    calculations using the method of \textcite{Kolosov1987}, with
    which the SAE \& TD-CIS results agree excellently. For helium and
    neon, we compare with experimental results (dashed, red)
    \cite{Rumble2021} and perturbation theory results (dot-dashed,
    grey). The static polarizability of neon as predicted by TD-CIS is
    slightly smaller (in magnitude) than the experimental result; the
    deviation can be explained by lack of correlation at the CIS
    level.}
\end{figure*}

The static polarizability is equivalent to the Stark shift, i.e.  how
much the ground state moves when applying a static electric field of
amplitude \(F\):
\begin{equation*}
  E(I) = E_0 -
  \frac{\alpha(0)}{2}F^2 -
  \frac{\beta(0)}{4}F^4 -
  \frac{\gamma(0)}{6}F^6 -
  \frac{\delta(0)}{8}F^8 - ....
\end{equation*}
For hydrogen, the first few values are \cite{Jentschura2001}
\begin{equation*}
  \begin{aligned}
    \alpha(0)
    &= \frac{\num{9}}{\num{2}},
    &
      \beta(0)
    &= \frac{\num{3555}}{16},
    \\
      \gamma(0)
    &= \frac{\num{2512779}\times6}{\num{512}},
    &
      \delta(0)
    &= \frac{\num{13012777803}}{\num{2048}}.
  \end{aligned}
\end{equation*}

We compute the static polarizability in a dynamic fashion \cite[see
section~6.1 of][]{Tannor2007} by applying a static field over a length
of time (i.e.\ sudden approximation), starting from an initial state
\(\Psi_i\) that is a linear combination of a few low-lying states
\begin{equation*}
  \ket{\Psi_i} = c_n\ket{n}.
\end{equation*}
At each time step, we compute the overlap of the wavefunction with
the initial state:
\begin{equation*}
  \begin{aligned}
    C(t)
    &=
      \braket{\Psi_i}{\Psi(t)} =
      \matrixel{\Psi_i}{\timeord\exp[-\im\int_0^t\diff{\tau}\Hamiltonian(\tau)]}{\Psi_i}
    \\
    &=
      \ce^{-\im E_m t} \abs{\braket{m}{\Psi_i}}^2,
  \end{aligned}
\end{equation*}
where in the last step we have used the fact that the field is static,
the Hamiltonian is time-independent, and hence we can trivially
rewrite the propagator on spectral form using the eigenstates
\(\ket{m}\) of the full Hamiltonian. From this we see that, as long as
\(\braket{m}{\Psi_i}\neq 0\), the correlation trace \(C(t)\) will contain
Fourier components at \(-E_m\); for moderate field strengths, the
field-free ground state is a large component of the ground state of
the full Hamiltonian, and we choose simply
\(\ket{\Psi_i}=\ket{\Psi_0}\). By repeating the calculation for different
field strengths \(F\), we can thus map \(\Delta E_m(F)\), which is shown
for hydrogen, helium, and neon in
Figure~\ref{fig:static-polarizability-hydrogen}.

% ** Dynamic polarizability
\subsection{Dynamic polarizability}
\begin{figure*}[htb]
  \centering
  \includegraphics{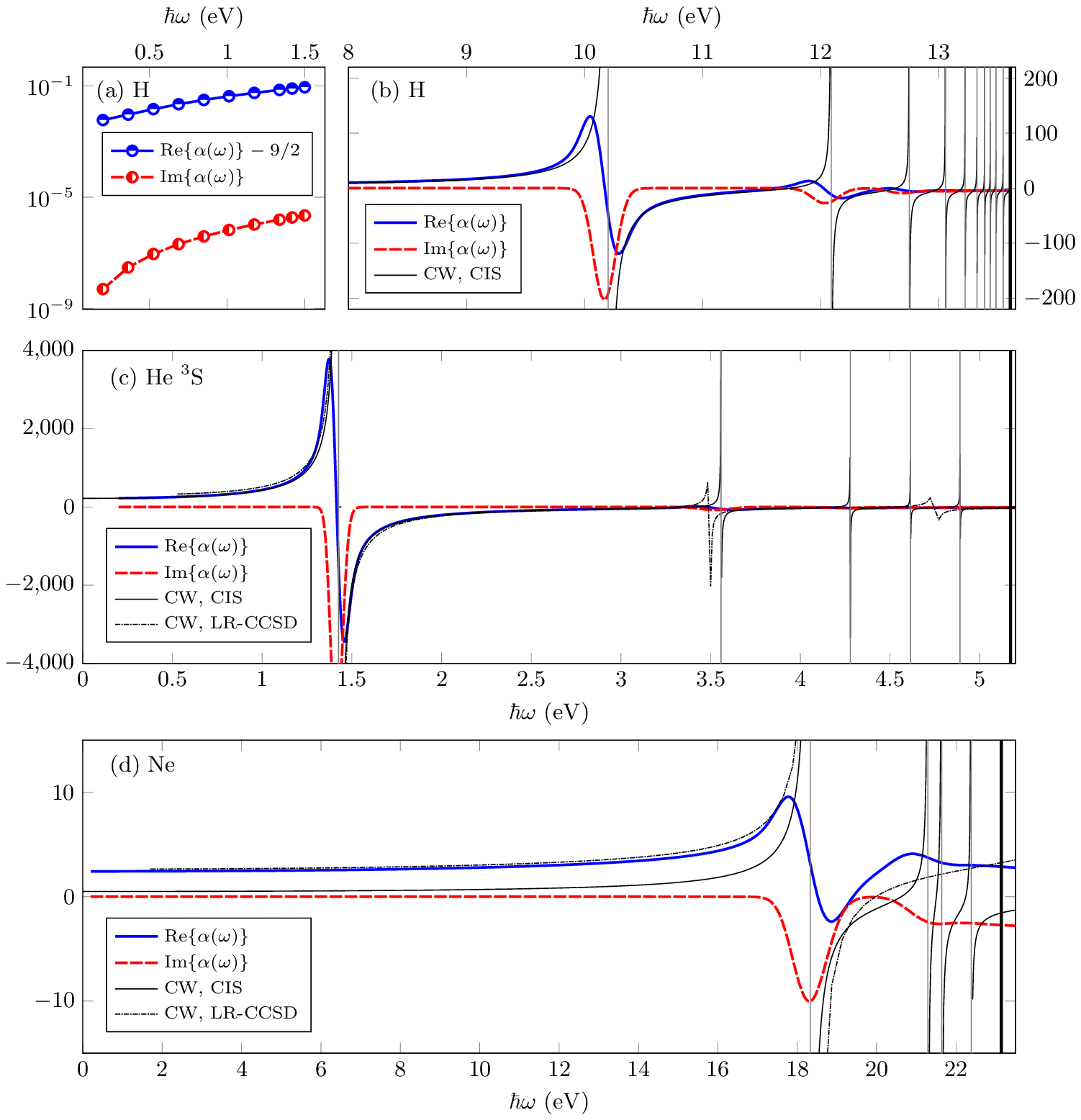}
  \caption{\label{fig:dynamic-polarizability}(a) Dynamic
    polarizability of the hydrogen ground state close to the adiabatic
    static-field limit; number of cycles \(\numcycles=2\),
    \(\ellmax=\conf{d}\). (b) Dynamic polarizability of the hydrogen
    ground state in the vicinity of the lowest resonances;
    \(\numcycles=10\), \(\ellmax=\conf{f}\). (c) Dynamic polarization
    of the \(\conf{1s\,2s}\,\term{3}S\) excited state of helium. The
    vertical lines indicate the positions of the CIS excited
    states. The first resonance above \SI{1}{\electronvolt} is the
    transition to \(\conf{1s\,2p}\,\term{3}{P}[1,2]\);
    \(\numcycles=4\), \(\ellmax=\conf{g}\). (d) Neon ground state;
    \(\numcycles=4\), \(\ellmax=\conf{g}\). The first resonance around
    \SI{18}{\electronvolt} is the transition to
    \(\conf{1s^2\,2s^2\,2p^5}\,(\term{2}{P}[3/2])\,\conf{3s}\), which
    experimentally is found around \SI{16}{\electronvolt}
    \cite{Kaufman1972JotOSoA,Saloman2004JoPaCRD}; the discrepancy is
    due to lack of correlation at the CIS level.}
\end{figure*}

The polarizability of a system relates the induced polarization to
the moments of the electric field:
\begin{equation*}
  \polarizationcomponent[^k](\omega_0;t) =
  \polarizability(\omega_0)
  \fieldcomponent[^k](t) +
  \hyperpolarizability[_{ij}^k](\omega_0)
  \fieldcomponent[^i](t)
  \fieldcomponent[^j](t) +
  ...
\end{equation*}
where \(\omega_0\) is the fundamental frequency of the electric
field \(\fieldamplitude[t]\), \(\polarizability\) is the (dynamic)
polarizability, and \(\hyperpolarizability\) is the
hyperpolarizability tensor.

To first order, we can compute \(\polarizability\) by driving the
system with a linearly polarized, nearly monochromatic pulse, and
dividing the Fourier transform of the induced dipole moment \(z(t)\)
at the driving frequency by the amplitude of the driving field:
\begin{equation*}
  \polarizability(\omega_0)=
  \frac{\hat{Z}(\omega_0)}{\hat{\fieldcomponent}(\omega_0)}.
\end{equation*}
We instead use the more numerically stable Mukamel expression
\cite[cf.\ e.g.\ Equation~(4.85) of][for the closely related medium
absorption]{Mukamel1995}
\begin{equation*}
  \expect{\polarizability}(\omega_0) \defd
  \frac{\displaystyle\int\diff{\omega'}\conj{\hat{F}_{\omega_0}}(\omega')\hat{Z}_{\omega_0}(\omega')}
  {\displaystyle\int\diff{\omega'}\conj{\hat{F}_{\omega_0}}(\omega')\hat{F}_{\omega_0}(\omega')},
\end{equation*}
which can be thought of as a weighted average, since the Fourier
transform of the driving field \(\hat{F}_{\omega_0}(\omega')\) is
peaked around \(\omega'=\omega_0\). For \(\omega_0\to0\), the
dynamic polarizabilities approach the static ones discussed in the
preceding section.

In Figure~\ref{fig:dynamic-polarizability}, dynamic polarizabilities for
the ground state of hydrogen, the triplet ground state of helium,
and the ground state of neon are presented. In the vicinity of
excited states accessible via one-photon absorption from the
initial state, the polarizabilities exhibit large resonances. For
comparison, we compute the dynamic polarizability of state \(i\)
for the case of continuous wave (CW), using the following formula
that involves the field-free energy differences and dipole moments
for excited states:
\begin{equation}
  \label{eqn:dynamic-polarizability-perturbation-theory}
  \alpha_{i;z}(\omega) =
  2\sumint_{k\neq i}
  \abs{\matrixel{i}{z}{k}}^2
  \frac{E_{ki}}{E_{ki}^2-\omega^2},
  \quad E_{ki}\defd E_k-E_i.
\end{equation}
Additionally, for \helium{} \term{3}{S} and \neon{} \term{1}{S}, we
compare with linear-response coupled-cluster singles+doubles
(LR-CCSD) calculations. For \helium{} LR-CCSD is essentially exact,
up to the quality of the basis set employed, in this case
\basisset{aug-cc-pV6Z} augmented with Rydberg-like Kaufmann basis
functions with \(n\leq8\) \cite{Kaufmann1989}. For \neon{} the
\basisset{aug-cc-pV6Z} basis set was used. Since these reference
calculations include more correlation than do the CIS results, the
peaks appear at slightly different energies. To enable a more
direct comparison, the curves have been shifted to approximately
align the first resonance; for \helium{} the shift is
\SI{0.28}{\electronvolt} and for \neon{} \SI{0.9}{\electronvolt}.

For all the dynamic polarizability calculations presented in
Figure~\ref{fig:dynamic-polarizability}, the truncated Gaussian
\eqref{eqn:truncated-gaussian} pulse was used, with a standard
deviation \(\sigma=2\pi\numcycles \omega^{-1}\), where \(\omega\)
is the driving pulse energy, and \(\numcycles\) the number of
cycles. \(\toff=3\sigma\), \(\tmax=5\sigma\), and the intensity was
\(I=\SI{e2}{\watt\per\centi\meter\squared}\). The dynamic
polarizabilities for finite pulses are in excellent agreement with
the perturbation theory predictions
\eqref{eqn:dynamic-polarizability-perturbation-theory}.

% ** Laser-induced resonant hole coupling
\subsection{Laser-induced resonant hole coupling}
\begin{figure}[htb]
  \centering
  \includegraphics{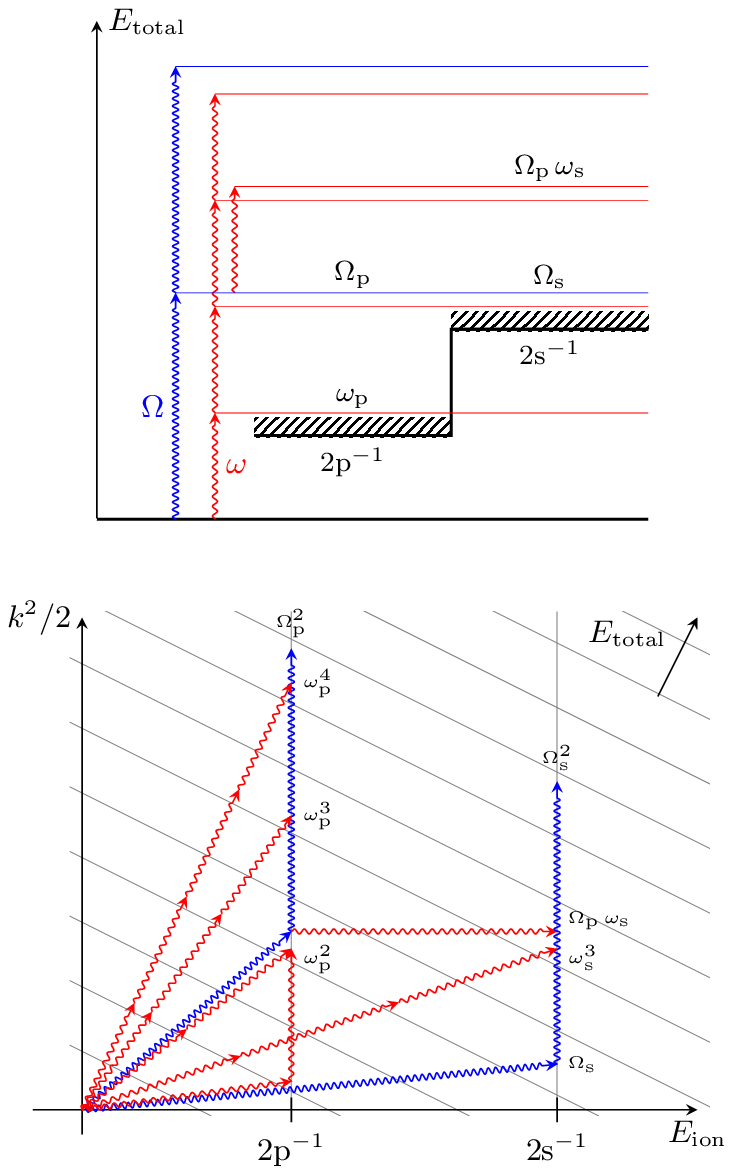}
  \caption{\label{fig:neon-ionization}Multichannel, multiphoton
    ionization of neon; in the CIS approximation, the \(\conf{2p}\)
    electron has an ionization potential of
    \(\sim\SI{23}{\electronvolt}\) and the \(\conf{2s}\) electron
    \(\sim\SI{53}{\electronvolt}\). The sketch illustrates weak-field
    ionization of neon into these two channels using a pump photon of
    energy \(\Omega\) and a probe photon of energy \(\omega\). The horizontal
    thin lines indicate where we expect to find photoelectron peaks,
    after absorbing various combinations of \(\Omega\) and \(\omega\). In the
    bottom panel, we show the same process in an energy-sharing
    diagram, where the abscissa shows energy of the ion, and the
    ordinate the kinetic energy of the photoelectron.}
\end{figure}
In multichannel ionization, where there is coupling between the
channels, it is vital that this coupling is properly accounted for
when solving the tSURFF equations of motion
\eqref{eqn:tsurff-solution}. An important example is the dipole
coupling between the holes in the residual ion, which has the
ability to move population from one channel to another, long after
the photoelectron has left the vicinity of the ion. If the
photoelectron wavepacket has already escaped the computational
domain through the matching sphere at \(\matchradius\), this effect
must be accounted for through \eqref{eqn:tsurff-solution},
otherwise there may be information missing from a ion-state
resolved spectrum. An analogous problem was studied for 1D neon by
\textcite{You2016}.

As a simple illustration of this mechanism, we consider the ionization
scheme illustrated in Figure~\ref{fig:neon-ionization}, where two
pulses with photon energies \(\Omega>\omega\) are used. We label the
photoelectron peaks according to the ionization pathways that led to
them, e.g.\ \(\peak{\Omega}{p}[2]\) corresponds to absorption of two
photons of energy \(\Omega\), leaving the ion in state
\(\slaterhole{\conf{2p}}\), whereas
\(\peak{\Omega}{p}\,\peak{\omega}{s}\) corresponds to ionization into
\(\slaterhole{\conf{2p}}\) with one photon of energy \(\Omega\), and then
channel coupling into \(\slaterhole{\conf{2s}}\) via one photon of
energy \(\omega\) (the label ordering indicates that \(\Omega\) in this case
arrives before \(\omega\)). In the complementary energy-sharing diagram
(cf.\ e.g.\ \cite{Smirnova2005}), the neutral atom is at zero energy,
i.e.\ the origin of the coordinate system. Diagonal lines mark
isolines of constant total energy of the system. Purely vertical
arrows indicate absorption of a photon by the photoelectron, and
similarly, purely horizontal arrows indicate absorption of a photon by
the ion. Ionization requires imparting energy on both the
photoelectron and the ion, which is why the corresponding arrows are
diagonal. We have chosen \(\omega=\Delta\ionpotential\), which is why the
photoelectron \(\peak{\Omega}{p}\,\peak{\omega}{s}\) (diagonal, followed by
horizontal arrow) will appear at the same kinetic energy as
\(\peak{\Omega}{p}\), and close to \(\peak{\omega}{s}[3]\).

\begin{figure*}[htb]
  \centering
  \includegraphics{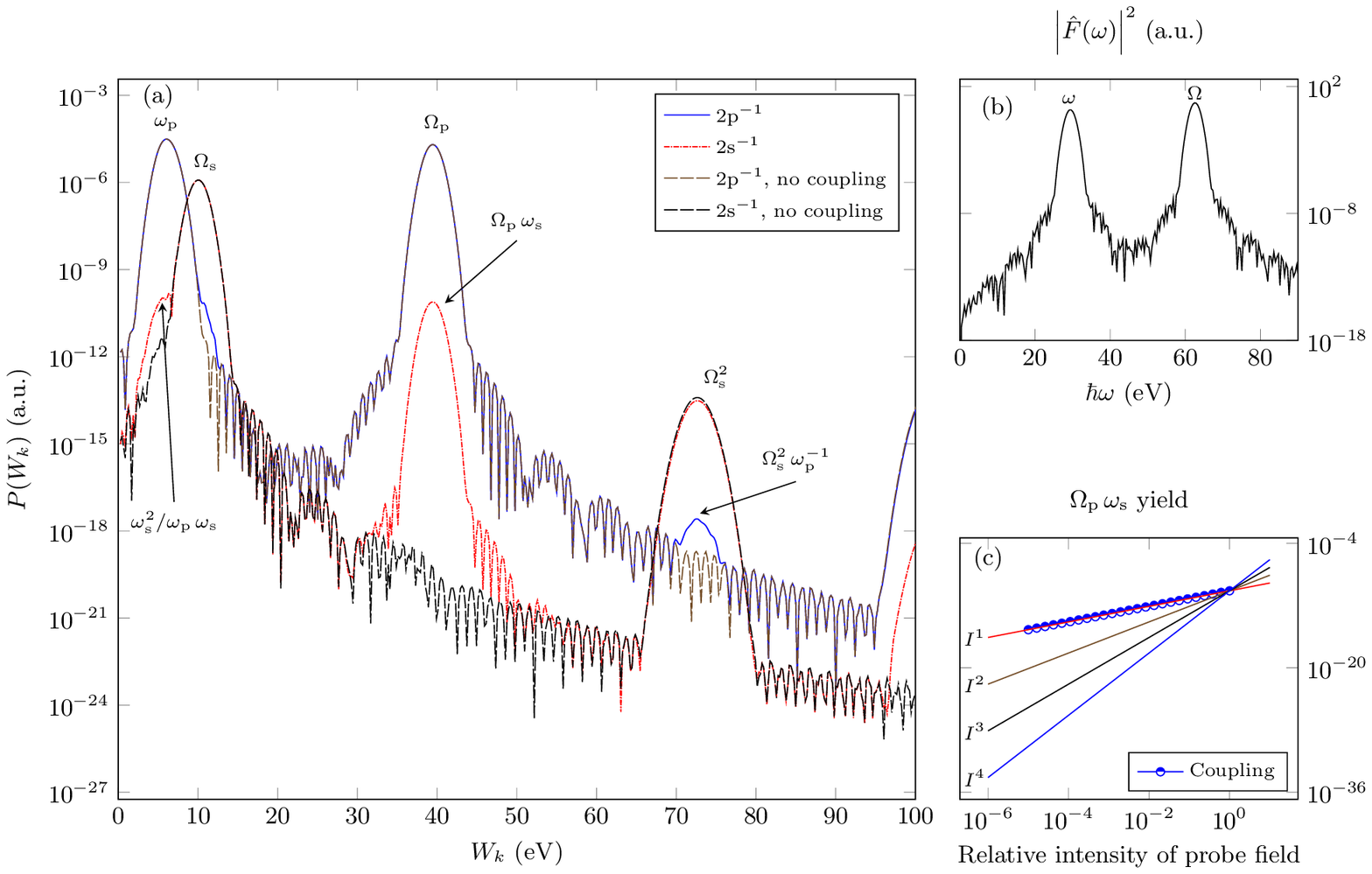}
  \caption{\label{fig:neon-spectrum}Ion-resolved spectra for neon.
    The pump pulse has an energy of
    \(\Omega=\SI{2.3}{\hartree}=\SI{62.6}{\electronvolt}\) and an intensity
    of \SI{3.51e10}{\watt\per\centi\meter\squared}. The probe pulse
    has an energy matching the difference in ionization potential,
    i.e.\ \(\omega\approx\SI{29.4}{\electronvolt}\) and an intensity of
    \SI{e10}{\watt\per\centi\meter\squared}. Both pulses have a
    duration of \SI{1}{\femto\second} (FWHM) and they are separated by
    approximately \SI{5}{\femto\second}.}
\end{figure*}

In Figure~\ref{fig:neon-spectrum}, the corresponding spectrum is
shown, resolved on final ion state and kinetic energy of the
photoelectron. The probe pulse with photon energy \(\omega\) is delayed
enough with respect to the pump pulse \(\Omega\), such that any electron
ejecta due to the latter has had time to leave the computational
domain entirely. The spectrum is computed both including the hole
couplings as well as neglecting them, the most obvious effect of which
is the disappearance of the peak
\(\peak{\Omega}{p}\,\peak{\omega}{s}\) at
\(\sim\SI{40}{\electronvolt}\) in the \(\slaterhole{\conf{2s}}\)
channel. To confirm that this peak arises mainly due to the proposed
ionization pathway, i.e.~\(\peak{\Omega}{p}\,\peak{\omega}{s}\), and not the
nearby possible pathways \(\peak{\omega}{p}[2]\,\peak{\omega}{s}\) and
\(\peak{\omega}{s}[3]\) (which are both third order in terms of
\(\omega\)), we repeat the calculation including the hole coupling, for a
range of intensities and observe the peak magnitude. Since the
magnitude of this peak varies linearly with the probe pulse intensity,
we conclude that it is due to absorption of one probe photon, and the
pathway \(\peak{\Omega}{p}\,\peak{\omega}{s}\) is the most likely one.

% ** Spin--orbit-split Fano resonances in neon
\subsection{Spin--orbit-split Fano resonances in neon}
To illustrate the power of iSURF \cite{Morales2016-isurf} in
resolving fine spectral details, we consider the ionization of neon
using a broadband attosecond pulse and study the autoionization due
to the Rydberg series
\(\conf{2s\,2p^6(\term{2}{S})}\,n\conf{p\;\term*{1}{P}[1]}\) that
is embedded in the
\(\conf{2s^2\,2p^5\;(\term*{2}{P}[\mathit{J}])}\,k\conf{s,d\;\term*{1}{P}[1]}\)
continuum (see Table~\ref{tab:neon-channels} for all possible pathways
accessible from the ground state, through absorption of 1--3
photons). The ionization is driven by a \SI{100}{\atto\second}
pulse of \SI{2}{\tera\watt\per\centi\meter\squared} centred at
\(\hbar\omega=\SI{1.124}{\hartree}\) (slightly above the
\(\slaterhole{\conf{2p}}\) threshold, but far below the
\(\slaterhole{\conf{2s}}\) threshold), the short pulse duration
corresponds to an energy bandwidth of
\(\sim\SI{11}{\electronvolt}\). We compare spectra from AE and RECP
calculations, where the RECP has been generated by
\textcite{Nicklass1995}.
\begin{table}[htb]
  \caption{\label{tab:neon-channels}Ionization channels accessible
    from the ground state of neon
    \(\conf{1s^2\,2s^2\,2p^6\;\term{1}S}\), through the absorption of
    at least \(q\) photons. The triplet terms are inaccessible in
    \(LS\) coupling, but spin--orbit interaction breaks this selection
    rule.}
  \centering
  \begin{ruledtabular}
    \begin{tabular}{rlll}
      Channel\Bstrut & \(\ell\) & \(q\) & Terms\\
      \hline
      \hline
      \(\conf{2s^2\,2p^5\;(\term*{2}{P}[\mathit{J}])}\,k\ell\)\Tstrut & \conf{s} & 1 & \(\term*{1}{P}, \term*{3}{P}\)\\
              & \conf{p} & 2 & \(\term{1}{S}, \term{1}{P}, \term{1}{D}, \term{3}{S}, \term{3}{P}, \term{3}{D}\)\\
              & \conf{d} & 1 & \(\term*{1}{P}, \term*{1}{D}, \term*{1}{F}, \term*{3}{P}, \term*{3}{D}, \term*{3}{F}\)\\
              & \conf{f}\Bstrut & 2 & \(\term{1}{D}, \term{1}{F}, \term{1}{G}, \term{3}{D}, \term{3}{F}, \term{3}{G}\)\\
      \hline
      \(\conf{2s\,2p^6(\term{2}{S})}\,n\ell\)\Tstrut & \conf{s} & 2 & \(\term{1}{S}, \term{3}{S}\)\\
              & \conf{p} & 1 & \(\term*{1}{P}, \term*{3}{P}\)\\
              & \conf{d} & 2 & \(\term{1}{D}, \term{3}{D}\)\\
              & \conf{f} & 3 & \(\term*{1}{F}, \term*{3}{F}\)
    \end{tabular}
  \end{ruledtabular}
\end{table}

Due to the spin--orbit interaction, the two allowed values of total
angular momentum for the intermediate term are \(J=3/2,1/2\), which
leads to a splitting of the autoionization resonances by
approximately \(\DEso=\SI{0.1}{\electronvolt}\); see
Table~\ref{tab:neon-energies}. The observed discrepancy between the
orbital energies of the AE calculation and the HF limit values
reported by \textcite{Fischer1977} is mostly due to the radial grid
employed \eqref{eqn:radial-grid}, which is not dense enough close
to the origin to accurately represent the \conf{1s} orbital. This
is also reflected in the total energy. However, since the \conf{1s}
orbital is the same in the initial and final states, this error
exactly cancels. The RECP results also differ from the reference
energies for the same reason; the \conf{1s} orbital is represented
using the RECP. We note that our spin--orbit splitting
(\(\SI{4.92}{\milli Ha}\approx\SI{0.134}{\electronvolt}\)) is
slightly closer to the experimental value \SI{0.097}{\electronvolt}
than what we are able to achieve with the \basisset{aug-cc-pVQZ}
basis set at the CIS level. Our splitting is also close to the
values computed using RCIS \cite{Zapata2022}, even if the absolute
orbital energies are slightly shifted.
\begin{table*}[htb]
  \caption{\label{tab:neon-energies}Energies (in Hartrees) of the
    occupied orbitals of the neon ground state, in the HF (DF\(^b\))
    approximation, for non-relativistic all-electron (AE), scalar
    relativistic effective-core potential (sRECP), and scalar+vector
    RECP calculations. In all cases, the radial grid extends to
    \(\rmax=\SI{400}{Bohr}\), with a non-uniform spacing of grid
    points following \eqref{eqn:radial-grid}.}
  \myruledtabular{l|S[table-format=4.8]S[table-format=4.6]|S[table-format=4.3e2]S[table-format=4.3e2]S[table-format=4.6]S[table-format=4.4e2]}{%
    Orbital\Tstrut & \mc{AE\(^a\)} & \mc{AE\(^*\)} & \mc{AE\(^b\)} & \mc{RECP\(^c\)} & \mc{sRECP\(^*\)} & \mc{RECP\(^*\)}\\
    \hline
    \(\conf{1s}\)\Tstrut & -32.7724455 & -31.5391 & -32.817475 &  &  & \\
    \(\conf{2s}\) & -1.93039095 & -1.9296 & -1.935847 & -1.931353 & -1.9497 & -1.9497\\
    \(\conf{2p}_{1/2}\) & -0.85040965 & -0.850289 & -0.852829 & -0.855553 & -0.845115 & -0.8484019\\
    \(\conf{2p}_{3/2}\) & '' & '' & -0.848267 & -0.848056 & '' & -0.8434788\\
    \(\DEso\)\Bstrut & 0 & 0 & 4.563e-3 & 7.497e-3 & 0 & 4.9231e-3\\
    \hline Total\Tstrut & -128.54710 & -126.283 & -128.691990 &
    -34.706149 & -34.6807 & -34.6807 }
  \begin{flushleft}
    \noindent
    \footnotesize{\(^a\) HF limit, \textcite{Fischer1977}}\\
    \footnotesize{\(^b\) Dirac--Fock values obtained from RCIS
      \cite{Zapata2022}}\\
    \footnotesize{\(^c\) Generated from the RECP by
      \textcite{Nicklass1995} using DIRAC19 \cite{DIRAC19} with a
      decontracted \basisset{aug-cc-pVQZ} basis set.\\
      \(^*\) this work}
  \end{flushleft}
\end{table*}

\begin{figure*}[htb]
  \centering
  \includegraphics{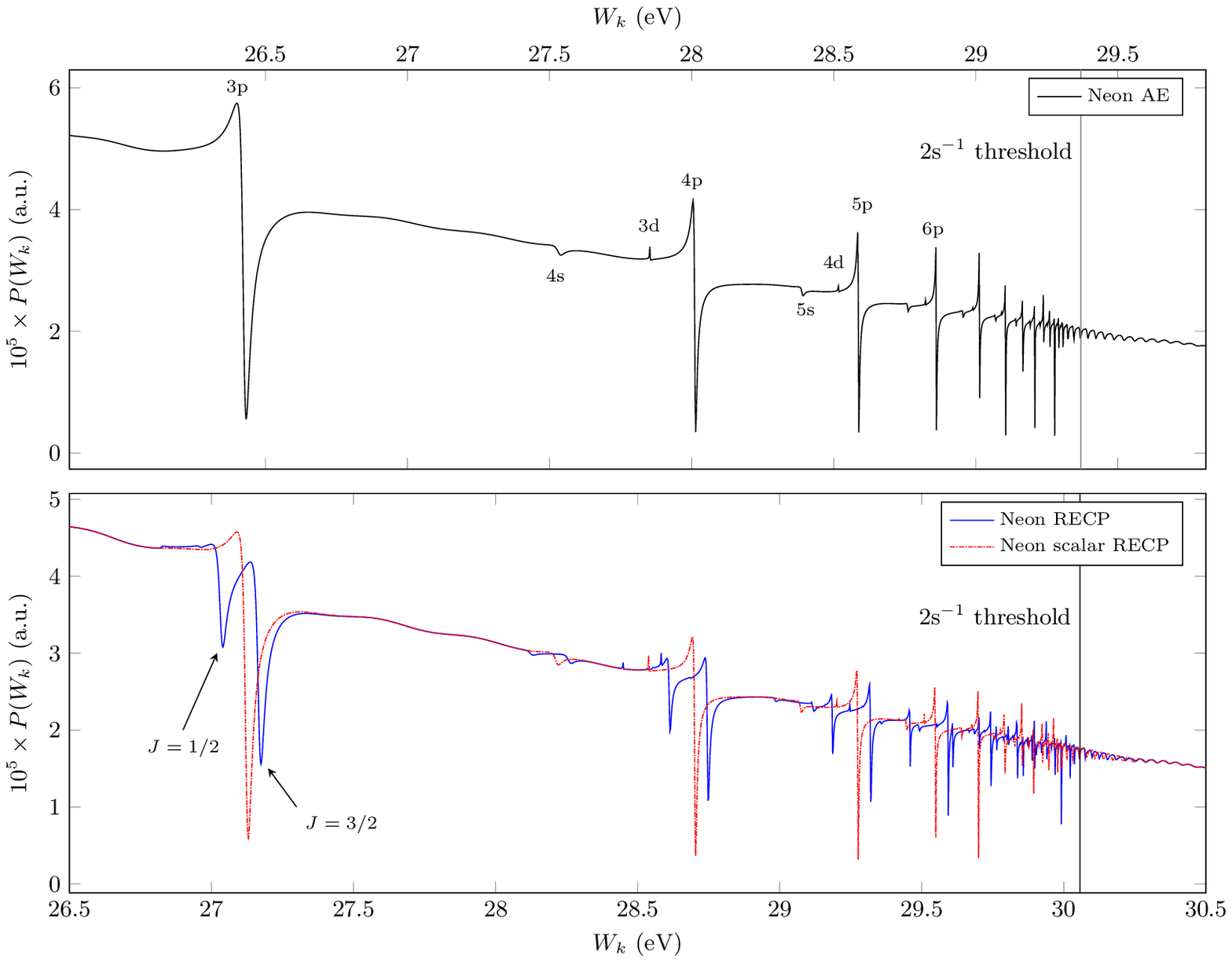}
  \caption{\label{fig:neon-fano}Photoionization spectrum of neon at
    the magic angle \(\theta=\arctan\sqrt{2}\) with respect to the
    polarization direction, in the vicinity of the Fano resonances
    converging to the \(\slaterhole{\conf{2s}}\) threshold, computed
    using various levels of theory; non-relativistic all-electron (AE)
    (\(\conf{1s}\) frozen, only \(\conf{2s,p}\) allowed to ionize) in
    the upper panel, and in the lower panel, a relativistic effective
    core potential (RECP) in solid blue, and finally its scalar
    counterpart in dot-dashed red. The panels have been aligned to the
    \(\slaterhole{\conf{2s}}\) threshold, see main text.}
\end{figure*}
An example photoelectron spectrum, calculated at various levels of
theory, is illustrated in Figure~\ref{fig:neon-fano}. The \(\DEso\)
splitting of the lines is clearly resolved; the corresponding
quantum beat period is
\(\qbperiod=2\pi/\DEso\sim\SI{43}{\femto\second}\). A normal
photoelectron spectrum calculation using tSURFF only would
necessitate post-propagation of the wavefunction, after the
ionizing electric field has turned off, by \emph{at least} \(\qbperiod\)
— in practice post-propagation on much longer timescales is
required. In contrast, using iSURF we can compute the spectrum
converged to infinite time directly after the end of the pulse, and
for the ultrashort pulse considered here, it is hardly a problem
too keep the whole wavefunction in the box, which enables us to use
the iSURF method with Coulomb asymptotics (iSURFC) which yields
essentially the exact spectrum, within our \emph{Ansatz}
\eqref{eqn:td-cis-ansatz}.

The large resonances are due to the intermediate states
\(\conf{2s\,2p^6(\term{2}{S})}\,n\conf{p\;\term*{1}{P}[1]}\) which
decay into
\(\conf{2s^2\,2p^5(\term*{2}{P}[\mathit{J}])}\,k\conf{s,d\;\term*{1}{P}[1]}\).
The precise location of these resonances differ between the AE and
RECP calculations, due to the difference in calculated orbital
energies (as seen in Table~\ref{tab:neon-energies}); the experimental
value for the \(\conf{3p}\) resonance is approximately
\SI{24}{\electronvolt} \cite{Codling1967,Saloman2004JoPaCRD}. The
centre-of-mass of the resonance lines agree between the scalar RECP
and the RECP, since the former is derived from the latter by
averaging the spin--orbit part of the potential. The splitting of
the lines are approximately \SI{0.1}{\electronvolt}, in agreement
with the spin--orbit splitting of the ion ground state. The
\(J=3/2\) lines appears at the higher kinetic energy of the
photoelectron, since the associated ionization potential is
lower. Its strengths are approximately double that of the \(J=1/2\)
line, due to the double number of available channels (four vs
two). The AE results appear consistently
\(\sim\SI{0.69}{\electronvolt}\) below the RECP results throughout
the Rydberg series, and the energy axis of the upper panel has been
shifted to reflect this. The location of the
\(\slaterhole{\conf{2s}}\) threshold has been computed in Koopman's
approximation, i.e.\ as the difference of the \(\conf{2p}\) and
\(\conf{2s}\) orbital energies; the experimental value is
\SI{26.9}{\electronvolt} \cite{Kramida2006TEPJD}, the large
discrepancy between theory and experiment is due to the lack of
relaxation of the remaining electrons at the CIS level.

Interspersed between the large resonances are two additional series,
one broader (corresponding to shorter autoionizing lifetimes) and the
other narrower (corresponding to longer lifetimes). These result from
the \(\conf{2s\,2p^6(\term{2}{S})}\,n\conf{s\;\term{1}{S}}\) and
\(\conf{2s\,2p^6(\term{2}{S})}\,n\conf{d\;\term{1}{D}}\) series,
respectively, which requires two photons to reach. The same
symmetries are found in the direct channel
\(\conf{2s^2\,2p^5(\term*{2}{P}[\mathit{J}])}\,k\ell\), for
\(\ell=\conf{p,f}\), which also require two photons, which explains why
these peaks are comparatively weaker. Although the
\(\conf{2s\,2p^6(\term{2}{S})}\,n\conf{s}\) series is unaffected by
spin--orbit interaction, the autoionization peaks are split due to the
splitting of the intermediate state
\(\conf{2s\,2p^6(\term{2}{S})}\,n'\conf{p}\).

% ** Spin-polarized photoelectrons in strong fields
\subsection{Spin-polarized photoelectrons in strong fields}
As a final example of the accuracy and power of the TD-CIS method,
we try to reproduce a recent experiment by \textcite{Trabert2018}, where
spin-polarized photoelectrons where produced by ionizing xenon
using intense circularly polarized light. In accordance with the
predictions by \textcite{Barth2013}, a strong connection between the
final ion state and the photoelectron spin is observed, leading to
high spin polarization of the ATI peaks. Additionally, the spin
polarization varies with the ATI peak order, since the tunnelling
conditions leading to a certain peak favour the co- and
counter-rotating spatial orbitals differently.

The radial grid of Equation~\eqref{eqn:radial-grid} used when solving
the HF problem consists of 40 points with the parameters given in
Table~\ref{tab:grid-parameters}, and \(\kinengmax=\SI{1}{\hartree}\)
\(\implies\rhomax=\SI{0.354}{Bohr}\) gives
\(\ionpotential[J=3/2]=\SI{0.446}{\hartree}\),
\(\ionpotential[J=1/2]=\SI{0.501}{\hartree}\). This results in a
spin--orbit splitting of
\(\DEso=\SI{54.5}{\milli\hartree}=\SI{1.48}{\electronvolt}\), which is
slightly larger than the experimental value
\(\DEso^{\textrm{exp}}\sim\SI{1.3}{\electronvolt}\). The time propagation
is performed on a larger grid of 217 points, where the spacing after
the 40th point is constant at \(\rho=\SI{0.341}{Bohr}\), extending to
\(\rmax\gtrsim\SI{70}{Bohr}\). The t+iSURF matching radius is at
\(\matchradius=\SI{41.86}{Bohr}\). Restricting excitation/ionization
to only occur from the \(5\conf{p}\) electrons (i.e.\ \num{6}
channels) and including all orbital angular momenta up to
\(\ellmax=15\) results in \num{3072} partial waves [see
Equation~\eqref{eqn:num-pws}].

The ionizing field is similar to the one used for the experiment; a
driving wavelength of \(\lambda=\SI{395}{\nano\meter}\), corresponding to a
pulse energy of \(\hbar\omega=\SI{3.14}{\electronvolt}\), an intensity of
\(I=\SI{60}{\tera\watt\per\centi\meter\squared}\), and (right-handed)
circular polarization. The plane of polarization is chosen to be the
\(x\)--\(y\) plane, such that the spin polarization along \(z\) is
non-zero. The pulse duration was however chosen as
\(\pulseduration=\SI{4}{\femto\second}\) instead of the
\SI{40}{\femto\second} of the experiment, since converging the spin
polarization for longer pulse durations becomes prohibitively
expensive; longer pulse durations necessitate larger orbital angular
momenta \(\ellmax\). Additionally, a pulse duration
\(\pulseduration=\SI{40}{\femto\second}\) would yield a pulse
bandwidth narrower than the typical error in transition energies due
to the CIS \emph{Ansatz}. A frequency scan would be required for
comparison with the experiment. When the energy errors are within the
bandwidth of the shorter pulse, the physical effect can robustly be
reproduced, without a parameter scan.

\begin{figure*}[htb]
  \centering
  \includegraphics{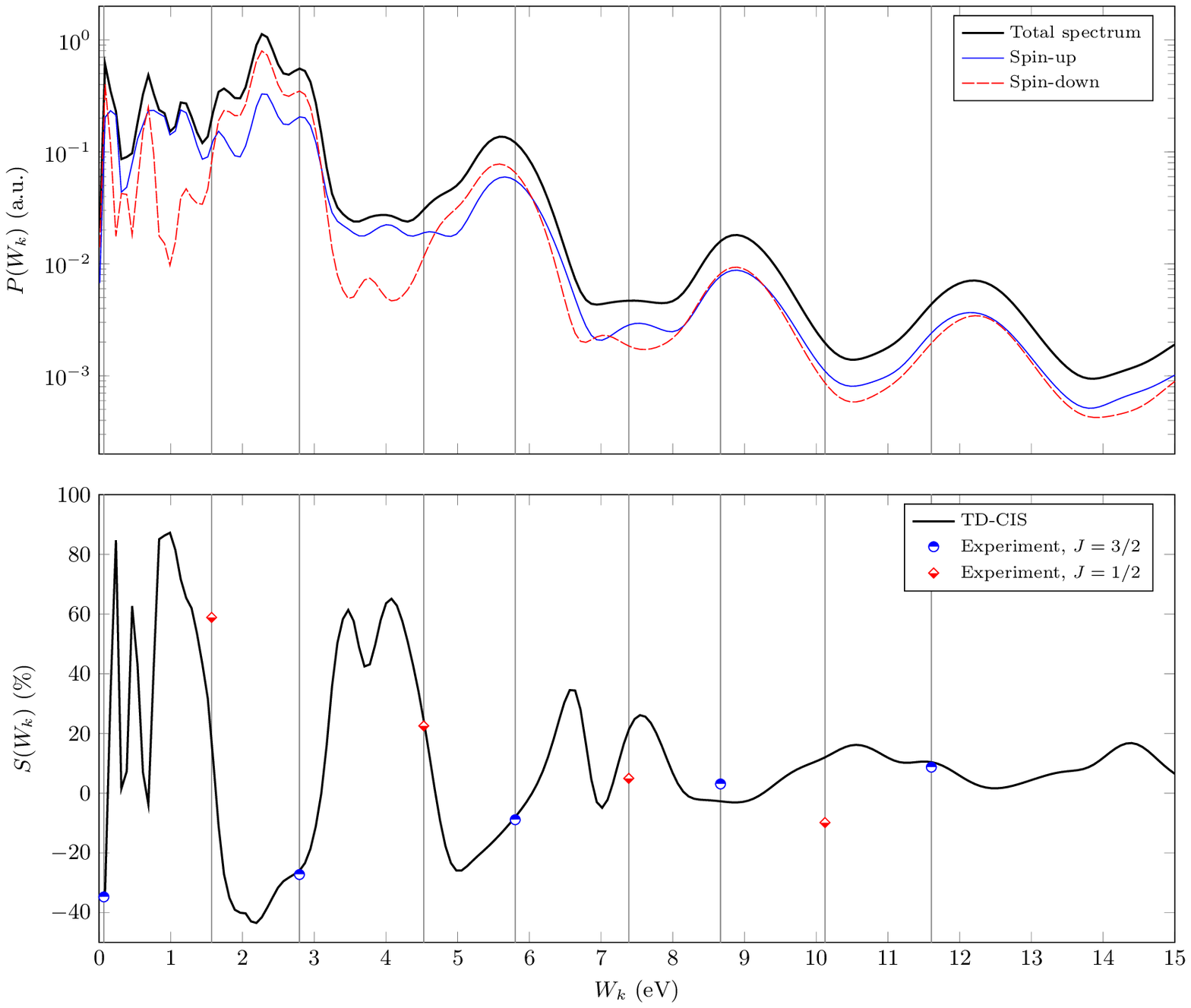}
  \caption{\label{fig:spin-polarization}Spin polarization of
    photoelectrons emitted during ionization of xenon using a
    circularly polarized field of
    \(I=\SI{60}{\tera\watt\per\centi\meter\squared}\) of
    \(\lambda=\SI{395}{\nano\meter}\). The ringing at low energies
    (\(\kineng<\SI{1}{\electronvolt}\)) is an artefact, due to the
    combination of a small computational box and a Volkov continuum
    \cite{Morales2016-isurf}; \(\matchradius\approx\SI{42}{Bohr}\).}
\end{figure*}

Finally, the photoelectron spectrum was resolved on a momentum grid,
with 200 points linearly spaced in energy from
\SIrange{0.01}{15}{\electronvolt}, 20 points along
\(\theta\in[0,\pi]\) and 41 points along \(\phi\in[0,2\pi]\). Volkov scattering states
(i.e.\ t+iSURFV) was employed, since the whole wavefunction could not
be kept within the computational box. The results are shown in
Figure~\ref{fig:spin-polarization}; the individual spectra for spin-up
and spin-down electrons (angularly integrated), respectively, as well
as the energy-resolved spin polarization, computed as
\begin{equation*}
  S[\si{\percent}] =
  100\frac{P_\alpha-P_\beta}{P_\alpha+P_\beta}.
\end{equation*}
Also shown are the experimental results by \textcite{Trabert2018}, with
which the theoretical spin polarization seems to be in satisfactory
agreement.

% * Conclusions
\section{Conclusions}
\label{sec:atomic-conclusions} We have described an efficient
propagator for the time-dependent configuration-interaction singles
\emph{Ansatz} specialized to the case of spherical symmetry, and
demonstrated its correctness with some simple examples, as well as its
ability to compute fine spectral features such as the spin--orbit
splitting of the Fano lines in photoionization of neon. We have
finally illustrated its capability to \emph{ab initio} faithfully
reproduce some of the latest experimental results pertaining to the
production of spin-polarized photoelectrons in strong-field processes,
an exciting area of research that is only going to grow in the near
future.
% * Acknowledgements
\begin{acknowledgments}
  It is a pleasure to thank Sheehan Olver for the collaboration on
  quasiarrays and Michael Spanner, Misha Yu.\ Ivanov, Morten
  Piibeleht, Pranav Singh, Rasmus Henningsson, Ken Schafer, and Sølve
  Selstø for illuminating discussions.

  SCM is supported by scholarship 185-608 from \emph{Olle Engkvists
    Stiftelse}. JMD acknowledges support from the \emph{Knut and Alice
    Wallenberg Foundation} (2017.0104 and 2019.0154), the
  \emph{Swedish Research Council} (2018-03845) and \emph{Olle
    Engkvists Stiftelse} (194-0734).
\end{acknowledgments}

% * Appendix
\appendix

% ** Quasimatrices
\section{Quasimatrices}
\label{sec:quasimatrices}
As stated in the introduction, the spin--angular degrees of freedom
are treated analytically using standard angular momentum
algebra. The radial degrees of freedom are described in the
language of \emph{quasimatrices} \cite{Olver2020}, which are objects
where the first dimension is formally continuous on the interval
\([a,b]\), and the second dimension discrete. Using a notation
reminiscent of that of Dirac, a quasimatrix can be written as
\begin{equation*}
  \quasimatrix{B} = \bmat{\quasifunction{1}&
    \quasifunction{2}&...&
    \quasifunction{n}},
\end{equation*}
where the columns \(\quasifunction{i}\) are functions on the
interval \([a,b]\), usually chosen to constitute a complete set on
this interval in the limit \(n\to\infty\). Any function may thus be
expanded as \(\quasifunction{f}=\quasimatrix{B}\vec{f}\), where
\(\vec{f}\) is the column vector of expansion
coefficients. Similarly, the one-dimensional time-independent
Schrödinger equation attains the familiar form of a generalized
eigenvalue equation
\begin{equation*}
  \begin{aligned}
    \Hamiltonian\quasifunction{\Psi}
    &=
      E\quasifunction{\Psi} \\
    \implies
    \adjoint{\quasimatrix{B}}
    \Hamiltonian
    \quasimatrix{B}
    \vec{c}
    &=
      \adjoint{\quasimatrix{B}}
      E
      \quasimatrix{B}
      \vec{c}
      \iff
      \Hammat
      \vec{c} =
      E
      \overlapmat
      \vec{c},
  \end{aligned}
\end{equation*}
where the matrix representation of the Hamiltonian is given by
\(\mat{H}_{ij}=\quasimatrixel*{i}{\Hamiltonian}{j}\), and the overlap
matrix \(\mat{S}_{ij}=\quasibraket{i}{j}\). In the case
\(\mat{S}_{ij}=\delta_{ij}\), we recover the standard eigenvalue
problem. The dual vectors \(\quasiadjfunction{i}\), and hence the dual
basis \(\adjoint{\quasimatrix{B}}\), are taken to be the complex
conjugates of \(\quasifunction{i}\) and \(\quasimatrix{B}\),
respectively. In non-Hermitian quantum mechanics, where the left and
right vectors in general do not coincide \cite{Moiseyev2011}, this is
strictly speaking an approximation.

The advantage of working with quasimatrices, instead of the matrix
representations of the operators directly, is that it becomes
easier to integrate different basis sets into the same code base;
to solve the HF problem, only the following operations need to be
implemented: basis function overlap \(\quasibraket{i}{j}\), scalar
functions (e.g.\ potentials) \(\quasimatrixel*{i}{\potop}{j}\),
derivatives \(\quasimatrixel*{i}{\partial^{(n)}}{j}\), for
\(n=1,2\), and mutual densities
\(\quasibraket{k}{i}\quasifunction{j}\) [corresponding to the
function product \(h(x)=f(x)g(x)\)].

The HF solver underpinning this TD-CIS implementation has been
implemented using quasimatrices \cite{ContinuumArrays0.9.0}, and
thus supports finite-differences of various kinds
\cite{Adler1984,Krause1999TJoPCA}, as well as finite-elements
discrete variable representation \cite{Rescigno2000}, and B-splines
\cite{Boor2001}. However, the time propagator component presently
requires diagonal overlap matrices \(\mat{S}_{ij}\sim\delta_{ij}\),
diagonal potential matrices, and tridiagonal derivative matrices,
for efficiency. An avenue of future improvement could be the
implementation of \emph{compact finite-differences}
\cite{Lele1992,Muller1999LP,Patchkovskii2016}, to increase the
spatial accuracy and potentially lowering the number of radial grid
points required.

% ** Givens rotations
\section{Givens rotations}
\label{sec:givens-rotations}
A common theme in the dipole propagators described in
section~\ref{sec:dipole-propagation} is the exponentiation of simple
\(2\times2\) systems, which can be computed using Givens rotations
\begin{equation*}
  \givens(s,c) \defd \bmat{c&s\\-\conj{s}&c}.
\end{equation*}
As an example, the complex-symmetric system
\begin{equation*}
  \mat{A} =
  -\im\bmat{0&a\\a&0}
\end{equation*}
can be exactly exponentiated as \(\exp(\mat{A})=
\givens(\cos a,-\im\sin a)\), but we instead opt for the
Crank--Nicolson approximation \cite{Muller1999LP}
\begin{equation*}
  \exp(\mat{A}) \approx
  \frac{1}{1+b^2}
  \bmat{1-b^2& -2\im b\\ -2\im b & 1-b^2} \equiv
  \frac{\givens(1-b^2,-2\im b)}{1+b^2}
\end{equation*}
(\(b \defd \frac{a}{2}\)), which is quicker to compute than the
trigonometric functions, while still being accurate enough for the
small rotation angles \(a\) considered.

% * Bibliography
\bibliography{tdcis-2-atoms}

\begin{thebibliography}{63}%
\makeatletter
\providecommand \@ifxundefined [1]{%
 \@ifx{#1\undefined}
}%
\providecommand \@ifnum [1]{%
 \ifnum #1\expandafter \@firstoftwo
 \else \expandafter \@secondoftwo
 \fi
}%
\providecommand \@ifx [1]{%
 \ifx #1\expandafter \@firstoftwo
 \else \expandafter \@secondoftwo
 \fi
}%
\providecommand \natexlab [1]{#1}%
\providecommand \enquote  [1]{``#1''}%
\providecommand \bibnamefont  [1]{#1}%
\providecommand \bibfnamefont [1]{#1}%
\providecommand \citenamefont [1]{#1}%
\providecommand \href@noop [0]{\@secondoftwo}%
\providecommand \href [0]{\begingroup \@sanitize@url \@href}%
\providecommand \@href[1]{\@@startlink{#1}\@@href}%
\providecommand \@@href[1]{\endgroup#1\@@endlink}%
\providecommand \@sanitize@url [0]{\catcode `\\12\catcode `\$12\catcode
  `\&12\catcode `\#12\catcode `\^12\catcode `\_12\catcode `\%12\relax}%
\providecommand \@@startlink[1]{}%
\providecommand \@@endlink[0]{}%
\providecommand \url  [0]{\begingroup\@sanitize@url \@url }%
\providecommand \@url [1]{\endgroup\@href {#1}{\urlprefix }}%
\providecommand \urlprefix  [0]{URL }%
\providecommand \Eprint [0]{\href }%
\providecommand \doibase [0]{https://doi.org/}%
\providecommand \selectlanguage [0]{\@gobble}%
\providecommand \bibinfo  [0]{\@secondoftwo}%
\providecommand \bibfield  [0]{\@secondoftwo}%
\providecommand \translation [1]{[#1]}%
\providecommand \BibitemOpen [0]{}%
\providecommand \bibitemStop [0]{}%
\providecommand \bibitemNoStop [0]{.\EOS\space}%
\providecommand \EOS [0]{\spacefactor3000\relax}%
\providecommand \BibitemShut  [1]{\csname bibitem#1\endcsname}%
\let\auto@bib@innerbib\@empty
%</preamble>
\bibitem [{\citenamefont {Carlström}\ \emph {et~al.}(2022)\citenamefont
  {Carlström}, \citenamefont {Spanner},\ and\ \citenamefont
  {Patchkovskii}}]{Carlstroem2022tdcisI}%
  \BibitemOpen
  \bibfield  {author} {\bibinfo {author} {\bibfnamefont {S.}~\bibnamefont
  {Carlström}}, \bibinfo {author} {\bibfnamefont {M.}~\bibnamefont
  {Spanner}},\ and\ \bibinfo {author} {\bibfnamefont {S.}~\bibnamefont
  {Patchkovskii}},\ }\bibfield  {title} {\bibinfo {title} {General
  time-dependent configuration-interaction singles {I}: The molecular case},\
  }\Eprint {https://arxiv.org/abs/2204.09966} {2204.09966}  (\bibinfo {year}
  {2022})\BibitemShut {NoStop}%
\bibitem [{\citenamefont {Varshalovich}(1988)}]{Varshalovich1988}%
  \BibitemOpen
  \bibfield  {author} {\bibinfo {author} {\bibfnamefont {D.~A.}\ \bibnamefont
  {Varshalovich}},\ }\href@noop {} {\emph {\bibinfo {title} {Quantum Theory of
  Angular Momentum: Irreducible Tensors, Spherical Harmonics, Vector Coupling
  Coefficients, 3nj Symbols}}}\ (\bibinfo  {publisher} {World Scientific Pub},\
  \bibinfo {address} {Singapore Teaneck, NJ, USA},\ \bibinfo {year}
  {1988})\BibitemShut {NoStop}%
\bibitem [{\citenamefont {Rohringer}\ \emph {et~al.}(2006)\citenamefont
  {Rohringer}, \citenamefont {Gordon},\ and\ \citenamefont
  {Santra}}]{Rohringer2006}%
  \BibitemOpen
  \bibfield  {author} {\bibinfo {author} {\bibfnamefont {N.}~\bibnamefont
  {Rohringer}}, \bibinfo {author} {\bibfnamefont {A.}~\bibnamefont {Gordon}},\
  and\ \bibinfo {author} {\bibfnamefont {R.}~\bibnamefont {Santra}},\
  }\bibfield  {title} {\bibinfo {title} {Configuration-interaction-based
  time-dependent orbital approach for \emph{Ab Initio} treatment of electronic
  dynamics in a strong optical laser field},\ }\href
  {https://doi.org/10.1103/physreva.74.043420} {\bibfield  {journal} {\bibinfo
  {journal} {Physical Review A}\ }\textbf {\bibinfo {volume} {74}},\ \bibinfo
  {pages} {043420} (\bibinfo {year} {2006})}\BibitemShut {NoStop}%
\bibitem [{\citenamefont {Rohringer}\ and\ \citenamefont
  {Santra}(2009)}]{Rohringer2009PRA}%
  \BibitemOpen
  \bibfield  {author} {\bibinfo {author} {\bibfnamefont {N.}~\bibnamefont
  {Rohringer}}\ and\ \bibinfo {author} {\bibfnamefont {R.}~\bibnamefont
  {Santra}},\ }\bibfield  {title} {\bibinfo {title} {Multichannel coherence in
  strong-field ionization},\ }\href
  {https://doi.org/10.1103/physreva.79.053402} {\bibfield  {journal} {\bibinfo
  {journal} {Physical Review A}\ }\textbf {\bibinfo {volume} {79}},\ \bibinfo
  {pages} {053402} (\bibinfo {year} {2009})}\BibitemShut {NoStop}%
\bibitem [{\citenamefont {Greenman}\ \emph {et~al.}(2010)\citenamefont
  {Greenman}, \citenamefont {Ho}, \citenamefont {Pabst}, \citenamefont
  {Kamarchik}, \citenamefont {Mazziotti},\ and\ \citenamefont
  {Santra}}]{Greenman2010PRA}%
  \BibitemOpen
  \bibfield  {author} {\bibinfo {author} {\bibfnamefont {L.}~\bibnamefont
  {Greenman}}, \bibinfo {author} {\bibfnamefont {P.~J.}\ \bibnamefont {Ho}},
  \bibinfo {author} {\bibfnamefont {S.}~\bibnamefont {Pabst}}, \bibinfo
  {author} {\bibfnamefont {E.}~\bibnamefont {Kamarchik}}, \bibinfo {author}
  {\bibfnamefont {D.~A.}\ \bibnamefont {Mazziotti}},\ and\ \bibinfo {author}
  {\bibfnamefont {R.}~\bibnamefont {Santra}},\ }\bibfield  {title} {\bibinfo
  {title} {Implementation of the time-dependent configuration-interaction
  singles method for atomic strong-field processes},\ }\href
  {https://doi.org/10.1103/physreva.82.023406} {\bibfield  {journal} {\bibinfo
  {journal} {Physical Review A}\ }\textbf {\bibinfo {volume} {82}},\ \bibinfo
  {pages} {023406} (\bibinfo {year} {2010})}\BibitemShut {NoStop}%
\bibitem [{\citenamefont {Harriman}(1978)}]{Harriman1978}%
  \BibitemOpen
  \bibfield  {author} {\bibinfo {author} {\bibfnamefont {J.}~\bibnamefont
  {Harriman}},\ }\href@noop {} {\emph {\bibinfo {title} {Theoretical
  foundations of electron spin resonance}}}\ (\bibinfo  {publisher} {Academic
  Press},\ \bibinfo {address} {New York},\ \bibinfo {year} {1978})\BibitemShut
  {NoStop}%
\bibitem [{\citenamefont {Dolg}\ and\ \citenamefont {Cao}(2011)}]{Dolg2011}%
  \BibitemOpen
  \bibfield  {author} {\bibinfo {author} {\bibfnamefont {M.}~\bibnamefont
  {Dolg}}\ and\ \bibinfo {author} {\bibfnamefont {X.}~\bibnamefont {Cao}},\
  }\bibfield  {title} {\bibinfo {title} {Relativistic pseudopotentials: Their
  development and scope of applications},\ }\href
  {https://doi.org/10.1021/cr2001383} {\bibfield  {journal} {\bibinfo
  {journal} {Chemical Reviews}\ }\textbf {\bibinfo {volume} {112}},\ \bibinfo
  {pages} {403} (\bibinfo {year} {2011})}\BibitemShut {NoStop}%
\bibitem [{\citenamefont {Zapata}\ \emph {et~al.}(2022)\citenamefont {Zapata},
  \citenamefont {Vinbladh}, \citenamefont {Ljungdahl}, \citenamefont
  {Lindroth},\ and\ \citenamefont {Dahlström}}]{Zapata2022}%
  \BibitemOpen
  \bibfield  {author} {\bibinfo {author} {\bibfnamefont {F.}~\bibnamefont
  {Zapata}}, \bibinfo {author} {\bibfnamefont {J.}~\bibnamefont {Vinbladh}},
  \bibinfo {author} {\bibfnamefont {A.}~\bibnamefont {Ljungdahl}}, \bibinfo
  {author} {\bibfnamefont {E.}~\bibnamefont {Lindroth}},\ and\ \bibinfo
  {author} {\bibfnamefont {J.~M.}\ \bibnamefont {Dahlström}},\ }\bibfield
  {title} {\bibinfo {title} {Relativistic time-dependent
  configuration-interaction singles method},\ }\href
  {https://doi.org/10.1103/physreva.105.012802} {\bibfield  {journal} {\bibinfo
   {journal} {Physical Review A}\ }\textbf {\bibinfo {volume} {105}},\ \bibinfo
  {pages} {012802} (\bibinfo {year} {2022})}\BibitemShut {NoStop}%
\bibitem [{\citenamefont {Indelicato}\ and\ \citenamefont
  {Desclaux}(1993)}]{Indelicato1993}%
  \BibitemOpen
  \bibfield  {author} {\bibinfo {author} {\bibfnamefont {P.}~\bibnamefont
  {Indelicato}}\ and\ \bibinfo {author} {\bibfnamefont {J.~P.}\ \bibnamefont
  {Desclaux}},\ }\bibfield  {title} {\bibinfo {title} {Projection operator in
  the multiconfiguration {D}irac--{F}ock method},\ }\href
  {https://doi.org/10.1088/0031-8949/1993/t46/015} {\bibfield  {journal}
  {\bibinfo  {journal} {Physica Scripta}\ }\textbf {\bibinfo {volume} {T46}},\
  \bibinfo {pages} {110} (\bibinfo {year} {1993})}\BibitemShut {NoStop}%
\bibitem [{\citenamefont {Indelicato}(1995)}]{Indelicato1995}%
  \BibitemOpen
  \bibfield  {author} {\bibinfo {author} {\bibfnamefont {P.}~\bibnamefont
  {Indelicato}},\ }\bibfield  {title} {\bibinfo {title} {Projection operators
  in multiconfiguration {D}irac--{F}ock calculations: Application to the ground
  state of heliumlike ions},\ }\href {https://doi.org/10.1103/physreva.51.1132}
  {\bibfield  {journal} {\bibinfo  {journal} {Physical Review A}\ }\textbf
  {\bibinfo {volume} {51}},\ \bibinfo {pages} {1132} (\bibinfo {year}
  {1995})}\BibitemShut {NoStop}%
\bibitem [{\citenamefont {Fischer}\ and\ \citenamefont
  {Zatsarinny}(2009)}]{Fischer2009}%
  \BibitemOpen
  \bibfield  {author} {\bibinfo {author} {\bibfnamefont {C.~F.}\ \bibnamefont
  {Fischer}}\ and\ \bibinfo {author} {\bibfnamefont {O.}~\bibnamefont
  {Zatsarinny}},\ }\bibfield  {title} {\bibinfo {title} {A {B}-spline
  {G}alerkin method for the {D}irac equation},\ }\href
  {https://doi.org/10.1016/j.cpc.2008.12.010} {\bibfield  {journal} {\bibinfo
  {journal} {Computer Physics Communications}\ }\textbf {\bibinfo {volume}
  {180}},\ \bibinfo {pages} {879} (\bibinfo {year} {2009})}\BibitemShut
  {NoStop}%
\bibitem [{\citenamefont {de~Walle}\ and\ \citenamefont
  {Blöchl}(1993)}]{Walle1993}%
  \BibitemOpen
  \bibfield  {author} {\bibinfo {author} {\bibfnamefont {C.~G.~V.}\
  \bibnamefont {de~Walle}}\ and\ \bibinfo {author} {\bibfnamefont {P.~E.}\
  \bibnamefont {Blöchl}},\ }\bibfield  {title} {\bibinfo {title}
  {First-principles calculations of hyperfine parameters},\ }\href
  {https://doi.org/10.1103/physrevb.47.4244} {\bibfield  {journal} {\bibinfo
  {journal} {Physical Review B}\ }\textbf {\bibinfo {volume} {47}},\ \bibinfo
  {pages} {4244} (\bibinfo {year} {1993})}\BibitemShut {NoStop}%
\bibitem [{\citenamefont {Pickard}\ and\ \citenamefont
  {Mauri}(2001)}]{Pickard2001}%
  \BibitemOpen
  \bibfield  {author} {\bibinfo {author} {\bibfnamefont {C.~J.}\ \bibnamefont
  {Pickard}}\ and\ \bibinfo {author} {\bibfnamefont {F.}~\bibnamefont
  {Mauri}},\ }\bibfield  {title} {\bibinfo {title} {All-electron magnetic
  response with pseudopotentials: {NMR} chemical shifts},\ }\href
  {https://doi.org/10.1103/physrevb.63.245101} {\bibfield  {journal} {\bibinfo
  {journal} {Physical Review B}\ }\textbf {\bibinfo {volume} {63}},\ \bibinfo
  {pages} {245101} (\bibinfo {year} {2001})}\BibitemShut {NoStop}%
\bibitem [{\citenamefont {Goddard}(2021)}]{Goddard2021}%
  \BibitemOpen
  \bibfield  {author} {\bibinfo {author} {\bibfnamefont {W.~A.}\ \bibnamefont
  {Goddard}},\ }\bibinfo {title} {\emph{Ab Initio} pseudopotentials (extending
  \emph{Ab Initio} {QM} throughout the periodic table)},\ in\ \href
  {https://doi.org/10.1007/978-3-030-18778-1_43} {\emph {\bibinfo {booktitle}
  {Computational Materials, Chemistry, and Biochemistry: From Bold Initiatives
  to the Last Mile}}},\ \bibinfo {series and number} {Computational Materials,
  Chemistry, and Biochemistry: From Bold Initiatives to the Last Mile}\
  (\bibinfo  {publisher} {Springer International Publishing},\ \bibinfo {year}
  {2021})\ pp.\ \bibinfo {pages} {1049--1053}\BibitemShut {NoStop}%
\bibitem [{\citenamefont {Beck}\ \emph {et~al.}(2009)\citenamefont {Beck},
  \citenamefont {Brozell}, \citenamefont {Blaudeau}, \citenamefont {Burggraf},\
  and\ \citenamefont {Pitzer}}]{Beck2009}%
  \BibitemOpen
  \bibfield  {author} {\bibinfo {author} {\bibfnamefont {E.~V.}\ \bibnamefont
  {Beck}}, \bibinfo {author} {\bibfnamefont {S.~R.}\ \bibnamefont {Brozell}},
  \bibinfo {author} {\bibfnamefont {J.-P.}\ \bibnamefont {Blaudeau}}, \bibinfo
  {author} {\bibfnamefont {L.~W.}\ \bibnamefont {Burggraf}},\ and\ \bibinfo
  {author} {\bibfnamefont {R.~M.}\ \bibnamefont {Pitzer}},\ }\bibfield  {title}
  {\bibinfo {title} {Assessment of the accuracy of shape-consistent
  relativistic effective core potentials using multireference spin-orbit
  configuration interaction singles and doubles calculations of the ground and
  low-lying excited states of {U}$^{4+}$ and {U}$^{5+}$},\ }\href
  {https://doi.org/10.1021/jp9049846} {\bibfield  {journal} {\bibinfo
  {journal} {The Journal of Physical Chemistry A}\ }\textbf {\bibinfo {volume}
  {113}},\ \bibinfo {pages} {12626} (\bibinfo {year} {2009})}\BibitemShut
  {NoStop}%
\bibitem [{\citenamefont {Odoh}\ and\ \citenamefont
  {Schreckenbach}(2009)}]{Odoh2009}%
  \BibitemOpen
  \bibfield  {author} {\bibinfo {author} {\bibfnamefont {S.~O.}\ \bibnamefont
  {Odoh}}\ and\ \bibinfo {author} {\bibfnamefont {G.}~\bibnamefont
  {Schreckenbach}},\ }\bibfield  {title} {\bibinfo {title} {Performance of
  relativistic effective core potentials in {DFT} calculations on actinide
  compounds},\ }\href {https://doi.org/10.1021/jp909576w} {\bibfield  {journal}
  {\bibinfo  {journal} {The Journal of Physical Chemistry A}\ }\textbf
  {\bibinfo {volume} {114}},\ \bibinfo {pages} {1957} (\bibinfo {year}
  {2009})}\BibitemShut {NoStop}%
\bibitem [{\citenamefont {Pabst}\ \emph {et~al.}(2012)\citenamefont {Pabst},
  \citenamefont {Sytcheva}, \citenamefont {Moulet}, \citenamefont {Wirth},
  \citenamefont {Goulielmakis},\ and\ \citenamefont {Santra}}]{Pabst2012}%
  \BibitemOpen
  \bibfield  {author} {\bibinfo {author} {\bibfnamefont {S.}~\bibnamefont
  {Pabst}}, \bibinfo {author} {\bibfnamefont {A.}~\bibnamefont {Sytcheva}},
  \bibinfo {author} {\bibfnamefont {A.}~\bibnamefont {Moulet}}, \bibinfo
  {author} {\bibfnamefont {A.}~\bibnamefont {Wirth}}, \bibinfo {author}
  {\bibfnamefont {E.}~\bibnamefont {Goulielmakis}},\ and\ \bibinfo {author}
  {\bibfnamefont {R.}~\bibnamefont {Santra}},\ }\bibfield  {title} {\bibinfo
  {title} {Theory of attosecond transient-absorption spectroscopy of krypton
  for overlapping pump and probe pulses},\ }\href
  {https://doi.org/10.1103/physreva.86.063411} {\bibfield  {journal} {\bibinfo
  {journal} {Physical Review A}\ }\textbf {\bibinfo {volume} {86}},\ \bibinfo
  {pages} {063411} (\bibinfo {year} {2012})}\BibitemShut {NoStop}%
\bibitem [{\citenamefont {Strang}(1968)}]{Strang1968}%
  \BibitemOpen
  \bibfield  {author} {\bibinfo {author} {\bibfnamefont {G.}~\bibnamefont
  {Strang}},\ }\bibfield  {title} {\bibinfo {title} {On the construction and
  comparison of difference schemes},\ }\href {https://doi.org/10.1137/0705041}
  {\bibfield  {journal} {\bibinfo  {journal} {SIAM Journal on Numerical
  Analysis}\ }\textbf {\bibinfo {volume} {5}},\ \bibinfo {pages} {506}
  (\bibinfo {year} {1968})}\BibitemShut {NoStop}%
\bibitem [{\citenamefont {Sato}\ \emph {et~al.}(2016)\citenamefont {Sato},
  \citenamefont {Ishikawa}, \citenamefont {Březinová}, \citenamefont
  {Lackner}, \citenamefont {Nagele},\ and\ \citenamefont
  {Burgdörfer}}]{Sato2016}%
  \BibitemOpen
  \bibfield  {author} {\bibinfo {author} {\bibfnamefont {T.}~\bibnamefont
  {Sato}}, \bibinfo {author} {\bibfnamefont {K.~L.}\ \bibnamefont {Ishikawa}},
  \bibinfo {author} {\bibfnamefont {I.}~\bibnamefont {Březinová}}, \bibinfo
  {author} {\bibfnamefont {F.}~\bibnamefont {Lackner}}, \bibinfo {author}
  {\bibfnamefont {S.}~\bibnamefont {Nagele}},\ and\ \bibinfo {author}
  {\bibfnamefont {J.}~\bibnamefont {Burgdörfer}},\ }\bibfield  {title}
  {\bibinfo {title} {Time-dependent complete-active-space self-consistent-field
  method for atoms: Application to high-order harmonic generation},\ }\href
  {https://doi.org/10.1103/physreva.94.023405} {\bibfield  {journal} {\bibinfo
  {journal} {Physical Review A}\ }\textbf {\bibinfo {volume} {94}},\ \bibinfo
  {pages} {023405} (\bibinfo {year} {2016})}\BibitemShut {NoStop}%
\bibitem [{\citenamefont {Teramura}\ \emph {et~al.}(2019)\citenamefont
  {Teramura}, \citenamefont {Sato},\ and\ \citenamefont
  {Ishikawa}}]{Teramura2019}%
  \BibitemOpen
  \bibfield  {author} {\bibinfo {author} {\bibfnamefont {T.}~\bibnamefont
  {Teramura}}, \bibinfo {author} {\bibfnamefont {T.}~\bibnamefont {Sato}},\
  and\ \bibinfo {author} {\bibfnamefont {K.~L.}\ \bibnamefont {Ishikawa}},\
  }\bibfield  {title} {\bibinfo {title} {Implementation of a gauge-invariant
  time-dependent configuration-interaction-singles method for three-dimensional
  atoms},\ }\href {https://doi.org/10.1103/physreva.100.043402} {\bibfield
  {journal} {\bibinfo  {journal} {Physical Review A}\ }\textbf {\bibinfo
  {volume} {100}},\ \bibinfo {pages} {043402} (\bibinfo {year}
  {2019})}\BibitemShut {NoStop}%
\bibitem [{\citenamefont {Hochbruck}\ and\ \citenamefont
  {Ostermann}(2010)}]{Hochbruck2010}%
  \BibitemOpen
  \bibfield  {author} {\bibinfo {author} {\bibfnamefont {M.}~\bibnamefont
  {Hochbruck}}\ and\ \bibinfo {author} {\bibfnamefont {A.}~\bibnamefont
  {Ostermann}},\ }\bibfield  {title} {\bibinfo {title} {Exponential
  integrators},\ }\href {https://doi.org/10.1017/s0962492910000048} {\bibfield
  {journal} {\bibinfo  {journal} {Acta Numerica}\ }\textbf {\bibinfo {volume}
  {19}},\ \bibinfo {pages} {209} (\bibinfo {year} {2010})}\BibitemShut
  {NoStop}%
\bibitem [{\citenamefont {Manolopoulos}(2002)}]{Manolopoulos2002}%
  \BibitemOpen
  \bibfield  {author} {\bibinfo {author} {\bibfnamefont {D.~E.}\ \bibnamefont
  {Manolopoulos}},\ }\bibfield  {title} {\bibinfo {title} {Derivation and
  reflection properties of a transmission-free absorbing potential},\ }\href
  {https://doi.org/10.1063/1.1517042} {\bibfield  {journal} {\bibinfo
  {journal} {J. Chem. Phys.}\ }\textbf {\bibinfo {volume} {117}},\ \bibinfo
  {pages} {9552} (\bibinfo {year} {2002})}\BibitemShut {NoStop}%
\bibitem [{\citenamefont {Riss}\ and\ \citenamefont {Meyer}(1993)}]{Riss1993}%
  \BibitemOpen
  \bibfield  {author} {\bibinfo {author} {\bibfnamefont {U.~V.}\ \bibnamefont
  {Riss}}\ and\ \bibinfo {author} {\bibfnamefont {H.-D.}\ \bibnamefont
  {Meyer}},\ }\bibfield  {title} {\bibinfo {title} {Calculation of resonance
  energies and widths using the complex absorbing potential method},\ }\href
  {https://doi.org/10.1088/0953-4075/26/23/021} {\bibfield  {journal} {\bibinfo
   {journal} {Journal of Physics B: Atomic, Molecular and Optical Physics}\
  }\textbf {\bibinfo {volume} {26}},\ \bibinfo {pages} {4503} (\bibinfo {year}
  {1993})}\BibitemShut {NoStop}%
\bibitem [{\citenamefont {Riss}\ and\ \citenamefont {Meyer}(1998)}]{Riss1998}%
  \BibitemOpen
  \bibfield  {author} {\bibinfo {author} {\bibfnamefont {U.~V.}\ \bibnamefont
  {Riss}}\ and\ \bibinfo {author} {\bibfnamefont {H.-D.}\ \bibnamefont
  {Meyer}},\ }\bibfield  {title} {\bibinfo {title} {The transformative complex
  absorbing potential method: a bridge between complex absorbing potentials and
  smooth exterior scaling},\ }\href
  {https://doi.org/10.1088/0953-4075/31/10/016} {\bibfield  {journal} {\bibinfo
   {journal} {Journal of Physics B: Atomic, Molecular and Optical Physics}\
  }\textbf {\bibinfo {volume} {31}},\ \bibinfo {pages} {2279} (\bibinfo {year}
  {1998})}\BibitemShut {NoStop}%
\bibitem [{\citenamefont {Moiseyev}(1998{\natexlab{a}})}]{Moiseyev1998a}%
  \BibitemOpen
  \bibfield  {author} {\bibinfo {author} {\bibfnamefont {N.}~\bibnamefont
  {Moiseyev}},\ }\bibfield  {title} {\bibinfo {title} {Derivations of universal
  exact complex absorption potentials by the generalized complex coordinate
  method},\ }\href {https://doi.org/10.1088/0953-4075/31/7/009} {\bibfield
  {journal} {\bibinfo  {journal} {Journal of Physics B: Atomic, Molecular and
  Optical Physics}\ }\textbf {\bibinfo {volume} {31}},\ \bibinfo {pages} {1431}
  (\bibinfo {year} {1998}{\natexlab{a}})}\BibitemShut {NoStop}%
\bibitem [{\citenamefont {Moiseyev}(1998{\natexlab{b}})}]{Moiseyev1998}%
  \BibitemOpen
  \bibfield  {author} {\bibinfo {author} {\bibfnamefont {N.}~\bibnamefont
  {Moiseyev}},\ }\bibfield  {title} {\bibinfo {title} {Quantum theory of
  resonances: calculating energies, widths and cross-sections by complex
  scaling},\ }\href {https://doi.org/10.1016/s0370-1573(98)00002-7} {\bibfield
  {journal} {\bibinfo  {journal} {Physics Reports}\ }\textbf {\bibinfo {volume}
  {302}},\ \bibinfo {pages} {212} (\bibinfo {year}
  {1998}{\natexlab{b}})}\BibitemShut {NoStop}%
\bibitem [{\citenamefont {Santra}\ and\ \citenamefont
  {Cederbaum}(2002)}]{Santra2002}%
  \BibitemOpen
  \bibfield  {author} {\bibinfo {author} {\bibfnamefont {R.}~\bibnamefont
  {Santra}}\ and\ \bibinfo {author} {\bibfnamefont {L.~S.}\ \bibnamefont
  {Cederbaum}},\ }\bibfield  {title} {\bibinfo {title} {Non-{Hermit}ian
  electronic theory and applications to clusters},\ }\href
  {https://doi.org/10.1016/s0370-1573(02)00143-6} {\bibfield  {journal}
  {\bibinfo  {journal} {Physics Reports}\ }\textbf {\bibinfo {volume} {368}},\
  \bibinfo {pages} {1} (\bibinfo {year} {2002})}\BibitemShut {NoStop}%
\bibitem [{\citenamefont {Scrinzi}(2010)}]{Scrinzi2010PRA}%
  \BibitemOpen
  \bibfield  {author} {\bibinfo {author} {\bibfnamefont {A.}~\bibnamefont
  {Scrinzi}},\ }\bibfield  {title} {\bibinfo {title} {Infinite-range exterior
  complex scaling as a perfect absorber in time-dependent problems},\ }\href
  {https://doi.org/10.1103/physreva.81.053845} {\bibfield  {journal} {\bibinfo
  {journal} {Physical Review A}\ }\textbf {\bibinfo {volume} {81}},\ \bibinfo
  {pages} {053845} (\bibinfo {year} {2010})}\BibitemShut {NoStop}%
\bibitem [{\citenamefont {Moiseyev}(2011)}]{Moiseyev2011}%
  \BibitemOpen
  \bibfield  {author} {\bibinfo {author} {\bibfnamefont {N.}~\bibnamefont
  {Moiseyev}},\ }\href@noop {} {\emph {\bibinfo {title} {Non-Hermitian Quantum
  Mechanics}}}\ (\bibinfo  {publisher} {Cambridge University Press},\ \bibinfo
  {address} {Cambridge New York},\ \bibinfo {year} {2011})\BibitemShut
  {NoStop}%
\bibitem [{\citenamefont {Froese~Fischer}(1977)}]{Fischer1977}%
  \BibitemOpen
  \bibfield  {author} {\bibinfo {author} {\bibfnamefont {C.}~\bibnamefont
  {Froese~Fischer}},\ }\href@noop {} {\emph {\bibinfo {title} {The
  {H}artree--{F}ock Method for Atoms: A Numerical Approach}}}\ (\bibinfo
  {publisher} {Wiley},\ \bibinfo {address} {New York},\ \bibinfo {year}
  {1977})\BibitemShut {NoStop}%
\bibitem [{\citenamefont {Fischer}\ and\ \citenamefont
  {Guo}(1990)}]{Fischer1990}%
  \BibitemOpen
  \bibfield  {author} {\bibinfo {author} {\bibfnamefont {C.~F.}\ \bibnamefont
  {Fischer}}\ and\ \bibinfo {author} {\bibfnamefont {W.}~\bibnamefont {Guo}},\
  }\bibfield  {title} {\bibinfo {title} {Spline algorithms for the
  {H}artree--{F}ock equation for the helium ground state},\ }\href
  {https://doi.org/10.1016/0021-9991(90)90176-2} {\bibfield  {journal}
  {\bibinfo  {journal} {Journal of Computational Physics}\ }\textbf {\bibinfo
  {volume} {90}},\ \bibinfo {pages} {486} (\bibinfo {year} {1990})}\BibitemShut
  {NoStop}%
\bibitem [{\citenamefont {McCurdy}\ \emph {et~al.}(2004)\citenamefont
  {McCurdy}, \citenamefont {Baertschy},\ and\ \citenamefont
  {Rescigno}}]{McCurdy2004}%
  \BibitemOpen
  \bibfield  {author} {\bibinfo {author} {\bibfnamefont {C.~W.}\ \bibnamefont
  {McCurdy}}, \bibinfo {author} {\bibfnamefont {M.}~\bibnamefont {Baertschy}},\
  and\ \bibinfo {author} {\bibfnamefont {T.~N.}\ \bibnamefont {Rescigno}},\
  }\bibfield  {title} {\bibinfo {title} {Solving the three-body {C}oulomb
  breakup problem using exterior complex scaling},\ }\href
  {https://doi.org/10.1088/0953-4075/37/17/r01} {\bibfield  {journal} {\bibinfo
   {journal} {Journal of Physics B: Atomic, Molecular and Optical Physics}\
  }\textbf {\bibinfo {volume} {37}},\ \bibinfo {pages} {R137} (\bibinfo {year}
  {2004})}\BibitemShut {NoStop}%
\bibitem [{\citenamefont {McCurdy}\ and\ \citenamefont
  {Martín}(2004)}]{McCurdy2004a}%
  \BibitemOpen
  \bibfield  {author} {\bibinfo {author} {\bibfnamefont {C.~W.}\ \bibnamefont
  {McCurdy}}\ and\ \bibinfo {author} {\bibfnamefont {F.}~\bibnamefont
  {Martín}},\ }\bibfield  {title} {\bibinfo {title} {Implementation of
  exterior complex scaling in {B}-splines to solve atomic and molecular
  collision problems},\ }\href {https://doi.org/10.1088/0953-4075/37/4/017}
  {\bibfield  {journal} {\bibinfo  {journal} {Journal of Physics B: Atomic,
  Molecular and Optical Physics}\ }\textbf {\bibinfo {volume} {37}},\ \bibinfo
  {pages} {917} (\bibinfo {year} {2004})}\BibitemShut {NoStop}%
\bibitem [{\citenamefont {Wolfsberg}(1955)}]{Wolfsberg1955}%
  \BibitemOpen
  \bibfield  {author} {\bibinfo {author} {\bibfnamefont {M.}~\bibnamefont
  {Wolfsberg}},\ }\bibfield  {title} {\bibinfo {title} {Dipole velocity and
  dipole length matrix elements in $\pi$ electron systems and configuration
  interaction},\ }\href {https://doi.org/10.1063/1.1742124} {\bibfield
  {journal} {\bibinfo  {journal} {The Journal of Chemical Physics}\ }\textbf
  {\bibinfo {volume} {23}},\ \bibinfo {pages} {793} (\bibinfo {year}
  {1955})}\BibitemShut {NoStop}%
\bibitem [{\citenamefont {Kobe}(1979)}]{Kobe1979}%
  \BibitemOpen
  \bibfield  {author} {\bibinfo {author} {\bibfnamefont {D.~H.}\ \bibnamefont
  {Kobe}},\ }\bibfield  {title} {\bibinfo {title} {Gauge-invariant resolution
  of the controversy over length versus velocity forms of the interaction with
  electric dipole radiation},\ }\href {https://doi.org/10.1103/physreva.19.205}
  {\bibfield  {journal} {\bibinfo  {journal} {Physical Review A}\ }\textbf
  {\bibinfo {volume} {19}},\ \bibinfo {pages} {205} (\bibinfo {year}
  {1979})}\BibitemShut {NoStop}%
\bibitem [{\citenamefont {Ishikawa}\ and\ \citenamefont
  {Sato}(2015)}]{Ishikawa2015}%
  \BibitemOpen
  \bibfield  {author} {\bibinfo {author} {\bibfnamefont {K.}~\bibnamefont
  {Ishikawa}}\ and\ \bibinfo {author} {\bibfnamefont {T.}~\bibnamefont
  {Sato}},\ }\bibfield  {title} {\bibinfo {title} {A review on \emph{Ab Initio}
  approaches for multielectron dynamics},\ }\href
  {https://doi.org/10.1109/jstqe.2015.2438827} {\bibfield  {journal} {\bibinfo
  {journal} {IEEE Journal of Selected Topics in Quantum Electronics}\ }\textbf
  {\bibinfo {volume} {21}},\ \bibinfo {pages} {8700916} (\bibinfo {year}
  {2015})}\BibitemShut {NoStop}%
\bibitem [{\citenamefont {Muller}(1999)}]{Muller1999LP}%
  \BibitemOpen
  \bibfield  {author} {\bibinfo {author} {\bibfnamefont {H.~G.}\ \bibnamefont
  {Muller}},\ }\bibfield  {title} {\bibinfo {title} {An efficient propagation
  scheme for the time-dependent {S}chrödinger equation in the velocity
  gauge},\ }\href
  {http://www.maik.ru/full/lasphys_archive/99/1/lasphys1_99p138full.pdf}
  {\bibfield  {journal} {\bibinfo  {journal} {Laser Physics}\ }\textbf
  {\bibinfo {volume} {9}},\ \bibinfo {pages} {138} (\bibinfo {year}
  {1999})}\BibitemShut {NoStop}%
\bibitem [{\citenamefont {Schafer}(2009)}]{Schafer2009}%
  \BibitemOpen
  \bibfield  {author} {\bibinfo {author} {\bibfnamefont {K.~J.}\ \bibnamefont
  {Schafer}},\ }\bibinfo {title} {Numerical methods in strong field physics}\
  (\bibinfo  {publisher} {Springer},\ \bibinfo {year} {2009})\ pp.\ \bibinfo
  {pages} {111--145}\BibitemShut {NoStop}%
\bibitem [{\citenamefont {Patchkovskii}\ and\ \citenamefont
  {Muller}(2016)}]{Patchkovskii2016}%
  \BibitemOpen
  \bibfield  {author} {\bibinfo {author} {\bibfnamefont {S.}~\bibnamefont
  {Patchkovskii}}\ and\ \bibinfo {author} {\bibfnamefont {H.}~\bibnamefont
  {Muller}},\ }\bibfield  {title} {\bibinfo {title} {Simple, accurate, and
  efficient implementation of 1-electron atomic time-dependent {S}chrödinger
  equation in spherical coordinates},\ }\href
  {https://doi.org/10.1016/j.cpc.2015.10.014} {\bibfield  {journal} {\bibinfo
  {journal} {Computer Physics Communications}\ }\textbf {\bibinfo {volume}
  {199}},\ \bibinfo {pages} {153} (\bibinfo {year} {2016})}\BibitemShut
  {NoStop}%
\bibitem [{\citenamefont {Krause}\ and\ \citenamefont
  {Schafer}(1999)}]{Krause1999TJoPCA}%
  \BibitemOpen
  \bibfield  {author} {\bibinfo {author} {\bibfnamefont {J.~L.}\ \bibnamefont
  {Krause}}\ and\ \bibinfo {author} {\bibfnamefont {K.~J.}\ \bibnamefont
  {Schafer}},\ }\bibfield  {title} {\bibinfo {title} {Control of {THz} emission
  from {S}tark wave packets},\ }\href {https://doi.org/10.1021/jp992144}
  {\bibfield  {journal} {\bibinfo  {journal} {The Journal of Physical Chemistry
  A}\ }\textbf {\bibinfo {volume} {103}},\ \bibinfo {pages} {10118} (\bibinfo
  {year} {1999})}\BibitemShut {NoStop}%
\bibitem [{\citenamefont {Kolosov}(1987)}]{Kolosov1987}%
  \BibitemOpen
  \bibfield  {author} {\bibinfo {author} {\bibfnamefont {V.~V.}\ \bibnamefont
  {Kolosov}},\ }\bibfield  {title} {\bibinfo {title} {A hydrogen atom in a
  strong electric field},\ }\href {https://doi.org/10.1088/0022-3700/20/11/008}
  {\bibfield  {journal} {\bibinfo  {journal} {Journal of Physics B: Atomic and
  Molecular Physics}\ }\textbf {\bibinfo {volume} {20}},\ \bibinfo {pages}
  {2359} (\bibinfo {year} {1987})}\BibitemShut {NoStop}%
\bibitem [{\citenamefont {Rumble}(2021)}]{Rumble2021}%
  \BibitemOpen
  \bibfield  {author} {\bibinfo {author} {\bibfnamefont {J.}~\bibnamefont
  {Rumble}},\ }\href@noop {} {\emph {\bibinfo {title} {CRC Handbook of
  Chemistry and Physics: A Ready-Reference Book of Chemical and Physical
  Data}}}\ (\bibinfo  {publisher} {CRC Press/Taylor \& Francis Group},\
  \bibinfo {address} {Boca Raton},\ \bibinfo {year} {2021})\BibitemShut
  {NoStop}%
\bibitem [{\citenamefont {Jentschura}(2001)}]{Jentschura2001}%
  \BibitemOpen
  \bibfield  {author} {\bibinfo {author} {\bibfnamefont {U.~D.}\ \bibnamefont
  {Jentschura}},\ }\bibfield  {title} {\bibinfo {title} {Resummation of the
  divergent perturbation series for a hydrogen atom in an electric field},\
  }\href {https://doi.org/10.1103/physreva.64.013403} {\bibfield  {journal}
  {\bibinfo  {journal} {Physical Review A}\ }\textbf {\bibinfo {volume} {64}},\
  \bibinfo {pages} {013403} (\bibinfo {year} {2001})}\BibitemShut {NoStop}%
\bibitem [{\citenamefont {Tannor}(2007)}]{Tannor2007}%
  \BibitemOpen
  \bibfield  {author} {\bibinfo {author} {\bibfnamefont {D.}~\bibnamefont
  {Tannor}},\ }\href@noop {} {\emph {\bibinfo {title} {Introduction to Quantum
  Mechanics: A Time-Dependent Perspective}}}\ (\bibinfo  {publisher}
  {University Science Books},\ \bibinfo {address} {Sausalito, Calif},\ \bibinfo
  {year} {2007})\BibitemShut {NoStop}%
\bibitem [{\citenamefont {Kaufman}\ and\ \citenamefont
  {Minnhagen}(1972)}]{Kaufman1972JotOSoA}%
  \BibitemOpen
  \bibfield  {author} {\bibinfo {author} {\bibfnamefont {V.}~\bibnamefont
  {Kaufman}}\ and\ \bibinfo {author} {\bibfnamefont {L.}~\bibnamefont
  {Minnhagen}},\ }\bibfield  {title} {\bibinfo {title} {Accurate ground-term
  combinations in {Ne I}},\ }\href {https://doi.org/10.1364/josa.62.000092}
  {\bibfield  {journal} {\bibinfo  {journal} {Journal of the Optical Society of
  America}\ }\textbf {\bibinfo {volume} {62}},\ \bibinfo {pages} {92} (\bibinfo
  {year} {1972})}\BibitemShut {NoStop}%
\bibitem [{\citenamefont {Saloman}(2004)}]{Saloman2004JoPaCRD}%
  \BibitemOpen
  \bibfield  {author} {\bibinfo {author} {\bibfnamefont {E.~B.}\ \bibnamefont
  {Saloman}},\ }\bibfield  {title} {\bibinfo {title} {Wavelengths, energy level
  classifications, and energy levels for the spectrum of neutral neon},\ }\href
  {https://doi.org/10.1063/1.1797771} {\bibfield  {journal} {\bibinfo
  {journal} {Journal of Physical and Chemical Reference Data}\ }\textbf
  {\bibinfo {volume} {33}},\ \bibinfo {pages} {1113} (\bibinfo {year}
  {2004})}\BibitemShut {NoStop}%
\bibitem [{\citenamefont {Mukamel}(1995)}]{Mukamel1995}%
  \BibitemOpen
  \bibfield  {author} {\bibinfo {author} {\bibfnamefont {S.}~\bibnamefont
  {Mukamel}},\ }\href@noop {} {\emph {\bibinfo {title} {Principles of Nonlinear
  Optical Spectroscopy}}}\ (\bibinfo  {publisher} {Oxford University Press},\
  \bibinfo {address} {New York},\ \bibinfo {year} {1995})\BibitemShut {NoStop}%
\bibitem [{\citenamefont {Kaufmann}\ \emph {et~al.}(1989)\citenamefont
  {Kaufmann}, \citenamefont {Baumeister},\ and\ \citenamefont
  {Jungen}}]{Kaufmann1989}%
  \BibitemOpen
  \bibfield  {author} {\bibinfo {author} {\bibfnamefont {K.}~\bibnamefont
  {Kaufmann}}, \bibinfo {author} {\bibfnamefont {W.}~\bibnamefont
  {Baumeister}},\ and\ \bibinfo {author} {\bibfnamefont {M.}~\bibnamefont
  {Jungen}},\ }\bibfield  {title} {\bibinfo {title} {Universal {G}aussian basis
  sets for an optimum representation of {R}ydberg and continuum
  wavefunctions},\ }\href {https://doi.org/10.1088/0953-4075/22/14/007}
  {\bibfield  {journal} {\bibinfo  {journal} {Journal of Physics B: Atomic,
  Molecular and Optical Physics}\ }\textbf {\bibinfo {volume} {22}},\ \bibinfo
  {pages} {2223} (\bibinfo {year} {1989})}\BibitemShut {NoStop}%
\bibitem [{\citenamefont {You}\ \emph {et~al.}(2016)\citenamefont {You},
  \citenamefont {Rohringer},\ and\ \citenamefont {Dahlström}}]{You2016}%
  \BibitemOpen
  \bibfield  {author} {\bibinfo {author} {\bibfnamefont {J.-A.}\ \bibnamefont
  {You}}, \bibinfo {author} {\bibfnamefont {N.}~\bibnamefont {Rohringer}},\
  and\ \bibinfo {author} {\bibfnamefont {J.~M.}\ \bibnamefont {Dahlström}},\
  }\bibfield  {title} {\bibinfo {title} {Attosecond photoionization dynamics
  with stimulated core-valence transitions},\ }\href
  {https://doi.org/10.1103/physreva.93.033413} {\bibfield  {journal} {\bibinfo
  {journal} {Physical Review A}\ }\textbf {\bibinfo {volume} {93}},\ \bibinfo
  {pages} {033413} (\bibinfo {year} {2016})}\BibitemShut {NoStop}%
\bibitem [{\citenamefont {Smirnova}\ \emph {et~al.}(2005)\citenamefont
  {Smirnova}, \citenamefont {Yakovlev},\ and\ \citenamefont
  {Ivanov}}]{Smirnova2005}%
  \BibitemOpen
  \bibfield  {author} {\bibinfo {author} {\bibfnamefont {O.}~\bibnamefont
  {Smirnova}}, \bibinfo {author} {\bibfnamefont {V.~S.}\ \bibnamefont
  {Yakovlev}},\ and\ \bibinfo {author} {\bibfnamefont {M.}~\bibnamefont
  {Ivanov}},\ }\bibfield  {title} {\bibinfo {title} {Use of electron
  correlation to make attosecond measurements without attosecond pulses},\
  }\href {https://doi.org/10.1103/physrevlett.94.213001} {\bibfield  {journal}
  {\bibinfo  {journal} {Physical Review Letters}\ }\textbf {\bibinfo {volume}
  {94}},\ \bibinfo {pages} {213001} (\bibinfo {year} {2005})}\BibitemShut
  {NoStop}%
\bibitem [{\citenamefont {Morales}\ \emph {et~al.}(2016)\citenamefont
  {Morales}, \citenamefont {Bredtmann},\ and\ \citenamefont
  {Patchkovskii}}]{Morales2016-isurf}%
  \BibitemOpen
  \bibfield  {author} {\bibinfo {author} {\bibfnamefont {F.}~\bibnamefont
  {Morales}}, \bibinfo {author} {\bibfnamefont {T.}~\bibnamefont {Bredtmann}},\
  and\ \bibinfo {author} {\bibfnamefont {S.}~\bibnamefont {Patchkovskii}},\
  }\bibfield  {title} {\bibinfo {title} {{iSURF}: a family of infinite-time
  surface flux methods},\ }\href
  {https://doi.org/10.1088/0953-4075/49/24/245001} {\bibfield  {journal}
  {\bibinfo  {journal} {Journal of Physics B: Atomic, Molecular and Optical
  Physics}\ }\textbf {\bibinfo {volume} {49}},\ \bibinfo {pages} {245001}
  (\bibinfo {year} {2016})}\BibitemShut {NoStop}%
\bibitem [{\citenamefont {Nicklass}\ \emph {et~al.}(1995)\citenamefont
  {Nicklass}, \citenamefont {Dolg}, \citenamefont {Stoll},\ and\ \citenamefont
  {Preuss}}]{Nicklass1995}%
  \BibitemOpen
  \bibfield  {author} {\bibinfo {author} {\bibfnamefont {A.}~\bibnamefont
  {Nicklass}}, \bibinfo {author} {\bibfnamefont {M.}~\bibnamefont {Dolg}},
  \bibinfo {author} {\bibfnamefont {H.}~\bibnamefont {Stoll}},\ and\ \bibinfo
  {author} {\bibfnamefont {H.}~\bibnamefont {Preuss}},\ }\bibfield  {title}
  {\bibinfo {title} {\emph{Ab Initio} energy‐adjusted pseudopotentials for
  the noble gases {Ne} through {Xe}: Calculation of atomic dipole and
  quadrupole polarizabilities},\ }\href {https://doi.org/10.1063/1.468948}
  {\bibfield  {journal} {\bibinfo  {journal} {The Journal of Chemical Physics}\
  }\textbf {\bibinfo {volume} {102}},\ \bibinfo {pages} {8942} (\bibinfo {year}
  {1995})}\BibitemShut {NoStop}%
\bibitem [{\citenamefont {Gomes}\ \emph {et~al.}(2019)\citenamefont {Gomes},
  \citenamefont {Saue}, \citenamefont {Visscher}, \citenamefont {Jensen},
  \citenamefont {Bast}, \citenamefont {Aucar}, \citenamefont {Bakken},
  \citenamefont {Dyall}, \citenamefont {S.~Dubillard}, \citenamefont {Eliav},
  \citenamefont {T.~Enevoldsen}, \citenamefont {Fleig}, \citenamefont
  {Fossgaard}, \citenamefont {L.~Halbert}, \citenamefont {Heimlich-Paris},
  \citenamefont {T.~Helgaker}, \citenamefont {Iliaš}, \citenamefont {Ch.
  R.~Jacob}, \citenamefont {Komorovský}, \citenamefont {O.~Kullie},
  \citenamefont {Larsen}, \citenamefont {Y.~S.~Lee}, \citenamefont {Nayak},
  \citenamefont {P.~Norman}, \citenamefont {Olsen}, \citenamefont {Olsen},
  \citenamefont {Y.~C.~Park}, \citenamefont {Pernpointner}, \citenamefont
  {R.~di Remigio}, \citenamefont {Sa{ł}ek}, \citenamefont
  {B.~Schimmelpfennig}, \citenamefont {Shee}, \citenamefont {Sikkema},
  \citenamefont {A.~J.~Thorvaldsen}, \citenamefont {van Stralen}, \citenamefont
  {M.~L.~Vidal}, \citenamefont {Visser}, \citenamefont {Winther},\ and\
  \citenamefont {Yamamoto}}]{DIRAC19}%
  \BibitemOpen
  \bibfield  {author} {\bibinfo {author} {\bibfnamefont {A.~S.~P.}\
  \bibnamefont {Gomes}}, \bibinfo {author} {\bibfnamefont {T.}~\bibnamefont
  {Saue}}, \bibinfo {author} {\bibfnamefont {L.}~\bibnamefont {Visscher}},
  \bibinfo {author} {\bibfnamefont {H.~J.~{\relax Aa}.}\ \bibnamefont
  {Jensen}}, \bibinfo {author} {\bibfnamefont {R.}~\bibnamefont {Bast}},
  \bibinfo {author} {\bibfnamefont {I.~A.}\ \bibnamefont {Aucar}}, \bibinfo
  {author} {\bibfnamefont {V.}~\bibnamefont {Bakken}}, \bibinfo {author}
  {\bibfnamefont {K.~G.}\ \bibnamefont {Dyall}}, \bibinfo {author}
  {\bibfnamefont {U.~E.}\ \bibnamefont {S.~Dubillard}}, \bibinfo {author}
  {\bibfnamefont {E.}~\bibnamefont {Eliav}}, \bibinfo {author} {\bibfnamefont
  {E.~F.}\ \bibnamefont {T.~Enevoldsen}}, \bibinfo {author} {\bibfnamefont
  {T.}~\bibnamefont {Fleig}}, \bibinfo {author} {\bibfnamefont
  {O.}~\bibnamefont {Fossgaard}}, \bibinfo {author} {\bibfnamefont {E.~D.~H.}\
  \bibnamefont {L.~Halbert}}, \bibinfo {author} {\bibfnamefont
  {B.}~\bibnamefont {Heimlich-Paris}}, \bibinfo {author} {\bibfnamefont
  {J.~H.}\ \bibnamefont {T.~Helgaker}}, \bibinfo {author} {\bibfnamefont
  {M.}~\bibnamefont {Iliaš}}, \bibinfo {author} {\bibfnamefont {S.~K.}\
  \bibnamefont {Ch. R.~Jacob}}, \bibinfo {author} {\bibfnamefont
  {S.}~\bibnamefont {Komorovský}}, \bibinfo {author} {\bibfnamefont
  {J.~K.~L.}\ \bibnamefont {O.~Kullie}}, \bibinfo {author} {\bibfnamefont
  {C.~V.}\ \bibnamefont {Larsen}}, \bibinfo {author} {\bibfnamefont {H.~S.~N.}\
  \bibnamefont {Y.~S.~Lee}}, \bibinfo {author} {\bibfnamefont {M.~K.}\
  \bibnamefont {Nayak}}, \bibinfo {author} {\bibfnamefont {G.~O.}\ \bibnamefont
  {P.~Norman}}, \bibinfo {author} {\bibfnamefont {J.}~\bibnamefont {Olsen}},
  \bibinfo {author} {\bibfnamefont {J.~M.~H.}\ \bibnamefont {Olsen}}, \bibinfo
  {author} {\bibfnamefont {J.~K.~P.}\ \bibnamefont {Y.~C.~Park}}, \bibinfo
  {author} {\bibfnamefont {M.}~\bibnamefont {Pernpointner}}, \bibinfo {author}
  {\bibfnamefont {K.~R.}\ \bibnamefont {R.~di Remigio}}, \bibinfo {author}
  {\bibfnamefont {P.}~\bibnamefont {Sa{ł}ek}}, \bibinfo {author}
  {\bibfnamefont {B.~S.}\ \bibnamefont {B.~Schimmelpfennig}}, \bibinfo {author}
  {\bibfnamefont {A.}~\bibnamefont {Shee}}, \bibinfo {author} {\bibfnamefont
  {J.}~\bibnamefont {Sikkema}}, \bibinfo {author} {\bibfnamefont {J.~T.}\
  \bibnamefont {A.~J.~Thorvaldsen}}, \bibinfo {author} {\bibfnamefont
  {J.}~\bibnamefont {van Stralen}}, \bibinfo {author} {\bibfnamefont {S.~V.}\
  \bibnamefont {M.~L.~Vidal}}, \bibinfo {author} {\bibfnamefont
  {O.}~\bibnamefont {Visser}}, \bibinfo {author} {\bibfnamefont
  {T.}~\bibnamefont {Winther}},\ and\ \bibinfo {author} {\bibfnamefont
  {S.}~\bibnamefont {Yamamoto}},\ }\href
  {https://doi.org/10.5281/zenodo.3572669} {\bibinfo {title} {{DIRAC19}, a
  relativistic \emph{Ab Initio} electronic structure program}} (\bibinfo {year}
  {2019})\BibitemShut {NoStop}%
\bibitem [{\citenamefont {Codling}\ \emph {et~al.}(1967)\citenamefont
  {Codling}, \citenamefont {Madden},\ and\ \citenamefont
  {Ederer}}]{Codling1967}%
  \BibitemOpen
  \bibfield  {author} {\bibinfo {author} {\bibfnamefont {K.}~\bibnamefont
  {Codling}}, \bibinfo {author} {\bibfnamefont {R.~P.}\ \bibnamefont
  {Madden}},\ and\ \bibinfo {author} {\bibfnamefont {D.~L.}\ \bibnamefont
  {Ederer}},\ }\bibfield  {title} {\bibinfo {title} {Resonances in the
  photo-ionization continuum of {Ne I} (20--150~{eV})},\ }\href
  {https://doi.org/10.1103/physrev.155.26} {\bibfield  {journal} {\bibinfo
  {journal} {Physical Review}\ }\textbf {\bibinfo {volume} {155}},\ \bibinfo
  {pages} {26} (\bibinfo {year} {1967})}\BibitemShut {NoStop}%
\bibitem [{\citenamefont {Kramida}\ and\ \citenamefont
  {Nave}(2006)}]{Kramida2006TEPJD}%
  \BibitemOpen
  \bibfield  {author} {\bibinfo {author} {\bibfnamefont {A.~E.}\ \bibnamefont
  {Kramida}}\ and\ \bibinfo {author} {\bibfnamefont {G.}~\bibnamefont {Nave}},\
  }\bibfield  {title} {\bibinfo {title} {The {Ne II} spectrum},\ }\href
  {https://doi.org/10.1140/epjd/e2006-00121-4} {\bibfield  {journal} {\bibinfo
  {journal} {The European Physical Journal D}\ }\textbf {\bibinfo {volume}
  {39}},\ \bibinfo {pages} {331} (\bibinfo {year} {2006})}\BibitemShut
  {NoStop}%
\bibitem [{\citenamefont {Trabert}\ \emph {et~al.}(2018)\citenamefont
  {Trabert}, \citenamefont {Hartung}, \citenamefont {Eckart}, \citenamefont
  {Trinter}, \citenamefont {Kalinin}, \citenamefont {Schöffler}, \citenamefont
  {Schmidt}, \citenamefont {Jahnke}, \citenamefont {Kunitski},\ and\
  \citenamefont {Dörner}}]{Trabert2018}%
  \BibitemOpen
  \bibfield  {author} {\bibinfo {author} {\bibfnamefont {D.}~\bibnamefont
  {Trabert}}, \bibinfo {author} {\bibfnamefont {A.}~\bibnamefont {Hartung}},
  \bibinfo {author} {\bibfnamefont {S.}~\bibnamefont {Eckart}}, \bibinfo
  {author} {\bibfnamefont {F.}~\bibnamefont {Trinter}}, \bibinfo {author}
  {\bibfnamefont {A.}~\bibnamefont {Kalinin}}, \bibinfo {author} {\bibfnamefont
  {M.}~\bibnamefont {Schöffler}}, \bibinfo {author} {\bibfnamefont {L.~P.~H.}\
  \bibnamefont {Schmidt}}, \bibinfo {author} {\bibfnamefont {T.}~\bibnamefont
  {Jahnke}}, \bibinfo {author} {\bibfnamefont {M.}~\bibnamefont {Kunitski}},\
  and\ \bibinfo {author} {\bibfnamefont {R.}~\bibnamefont {Dörner}},\
  }\bibfield  {title} {\bibinfo {title} {Spin and angular momentum in
  strong-field ionization},\ }\href
  {https://doi.org/10.1103/physrevlett.120.043202} {\bibfield  {journal}
  {\bibinfo  {journal} {Physical Review Letters}\ }\textbf {\bibinfo {volume}
  {120}},\ \bibinfo {pages} {043202} (\bibinfo {year} {2018})}\BibitemShut
  {NoStop}%
\bibitem [{\citenamefont {Barth}\ and\ \citenamefont
  {Smirnova}(2013)}]{Barth2013}%
  \BibitemOpen
  \bibfield  {author} {\bibinfo {author} {\bibfnamefont {I.}~\bibnamefont
  {Barth}}\ and\ \bibinfo {author} {\bibfnamefont {O.}~\bibnamefont
  {Smirnova}},\ }\bibfield  {title} {\bibinfo {title} {Spin-polarized electrons
  produced by strong-field ionization},\ }\href
  {https://doi.org/10.1103/physreva.88.013401} {\bibfield  {journal} {\bibinfo
  {journal} {Physical Review A}\ }\textbf {\bibinfo {volume} {88}},\ \bibinfo
  {pages} {013401} (\bibinfo {year} {2013})}\BibitemShut {NoStop}%
\bibitem [{\citenamefont {Olver}\ \emph {et~al.}(2020)\citenamefont {Olver},
  \citenamefont {Slevinsky},\ and\ \citenamefont {Townsend}}]{Olver2020}%
  \BibitemOpen
  \bibfield  {author} {\bibinfo {author} {\bibfnamefont {S.}~\bibnamefont
  {Olver}}, \bibinfo {author} {\bibfnamefont {R.~M.}\ \bibnamefont
  {Slevinsky}},\ and\ \bibinfo {author} {\bibfnamefont {A.}~\bibnamefont
  {Townsend}},\ }\bibfield  {title} {\bibinfo {title} {Fast algorithms using
  orthogonal polynomials},\ }\href {https://doi.org/10.1017/s0962492920000045}
  {\bibfield  {journal} {\bibinfo  {journal} {Acta Numerica}\ }\textbf
  {\bibinfo {volume} {29}},\ \bibinfo {pages} {573} (\bibinfo {year}
  {2020})}\BibitemShut {NoStop}%
\bibitem [{\citenamefont {Olver}\ \emph {et~al.}(2021)\citenamefont {Olver},
  \citenamefont {Carlström},\ and\ \citenamefont
  {Gutleb}}]{ContinuumArrays0.9.0}%
  \BibitemOpen
  \bibfield  {author} {\bibinfo {author} {\bibfnamefont {S.}~\bibnamefont
  {Olver}}, \bibinfo {author} {\bibfnamefont {S.}~\bibnamefont {Carlström}},\
  and\ \bibinfo {author} {\bibfnamefont {T.~S.}\ \bibnamefont {Gutleb}},\
  }\href {https://doi.org/10.5281/ZENODO.5151531} {\bibinfo {title}
  {{J}ulia{A}pproximation/{C}ontinuum{A}rrays.jl: v0.9.0}} (\bibinfo {year}
  {2021})\BibitemShut {NoStop}%
\bibitem [{\citenamefont {Adler}\ and\ \citenamefont
  {Piran}(1984)}]{Adler1984}%
  \BibitemOpen
  \bibfield  {author} {\bibinfo {author} {\bibfnamefont {S.~L.}\ \bibnamefont
  {Adler}}\ and\ \bibinfo {author} {\bibfnamefont {T.}~\bibnamefont {Piran}},\
  }\bibfield  {title} {\bibinfo {title} {Relaxation methods for gauge field
  equilibrium equations},\ }\href {https://doi.org/10.1103/revmodphys.56.1}
  {\bibfield  {journal} {\bibinfo  {journal} {Reviews of Modern Physics}\
  }\textbf {\bibinfo {volume} {56}},\ \bibinfo {pages} {1} (\bibinfo {year}
  {1984})}\BibitemShut {NoStop}%
\bibitem [{\citenamefont {Rescigno}\ and\ \citenamefont
  {McCurdy}(2000)}]{Rescigno2000}%
  \BibitemOpen
  \bibfield  {author} {\bibinfo {author} {\bibfnamefont {T.~N.}\ \bibnamefont
  {Rescigno}}\ and\ \bibinfo {author} {\bibfnamefont {C.~W.}\ \bibnamefont
  {McCurdy}},\ }\bibfield  {title} {\bibinfo {title} {Numerical grid methods
  for quantum-mechanical scattering problems},\ }\href
  {https://doi.org/10.1103/physreva.62.032706} {\bibfield  {journal} {\bibinfo
  {journal} {Physical Review A}\ }\textbf {\bibinfo {volume} {62}},\ \bibinfo
  {pages} {032706} (\bibinfo {year} {2000})}\BibitemShut {NoStop}%
\bibitem [{\citenamefont {{de Boor}}(2001)}]{Boor2001}%
  \BibitemOpen
  \bibfield  {author} {\bibinfo {author} {\bibfnamefont {C.}~\bibnamefont {{de
  Boor}}},\ }\href@noop {} {\emph {\bibinfo {title} {A Practical Guide to
  Splines: With 32 Figures}}}\ (\bibinfo  {publisher} {Springer},\ \bibinfo
  {address} {New York},\ \bibinfo {year} {2001})\BibitemShut {NoStop}%
\bibitem [{\citenamefont {Lele}(1992)}]{Lele1992}%
  \BibitemOpen
  \bibfield  {author} {\bibinfo {author} {\bibfnamefont {S.~K.}\ \bibnamefont
  {Lele}},\ }\bibfield  {title} {\bibinfo {title} {Compact finite difference
  schemes with spectral-like resolution},\ }\href
  {https://doi.org/10.1016/0021-9991(92)90324-r} {\bibfield  {journal}
  {\bibinfo  {journal} {Journal of Computational Physics}\ }\textbf {\bibinfo
  {volume} {103}},\ \bibinfo {pages} {16} (\bibinfo {year} {1992})}\BibitemShut
  {NoStop}%
\end{thebibliography}%

% * Fin

\end{document}